\documentclass[pre,onecolumn,preprintnumbers,floatfix,amssymb,amsmath]{revtex4}
\usepackage{graphicx}
\usepackage{subfigure}

\begin{document}

\title{Contact angle hysteresis effects on a drop sitting on an incline plane}
\author{ Pablo D. Ravazzoli, Ingrith Cuellar, Alejandro G. Gonz\'alez and Javier A. Diez}
\affiliation{Instituto de F\'{\i}sica Arroyo Seco, Universidad Nacional del
Centro de la Provincia de Buenos Aires, and CIFICEN-CONICET-CICPBA, Pinto 399,
7000, Tandil, Argentina}

\begin{abstract}
We study the contact angle hysteresis and morphology changes of a liquid drop
sitting on a solid substrate inclined with respect to the horizontal at an
angle $\alpha$. This one is always smaller than the critical one, $\alpha_{crit}$,
above which the drop would start to slide down. The hysteresis cycle is
performed for positive and negative $\alpha$'s ($|\alpha|<\alpha_{crit}$), and a
complete study of the changes in contact angles, free surface and footprint
shape is carried out. The drop shape is analyzed in terms of a solution of the
equilibrium pressure equation within and out of the long--wave model 
(lubrication approximation). Within this approximation, we obtain a truncated analytical
solution that describes the static drop shapes, that is successfully compared with experimental data. This solution is of practical interest since it allows for a complete description of all the drop features, such as its footprint shape or contact angle distribution around the drop periphery, starting 
from a very small set of relatively easy to measure drop parameters.
\end{abstract}

\maketitle
\section{Introduction}
\label{sec:intro}
The problem of a sessile drop sitting on an incline has been the subject of
numerous investigations in the last three decades. Some of them have been
concerned with the relationship between the maximum plane inclination above
which the drop slides down, $\alpha_{crit}$, and the parameters characterizing
the initial conditions~\cite{dussan_jfm85,quere_lang98,pierce_csA08,janardan_csA14}. Many
studies aim to describe the retention forces needed to achieve this critical
inclination angle~\cite{extrand_jcis90,extrand_jcis95,krasovitski_lang05,elsherbini_jcis06,gao_nphys18},
while other related works focus on the dynamics of droplets sliding down an
incline~\cite{legrand_jfm05,anapragada_ijhmt11}.

\begin{figure}[htb]
\begin{center}
\includegraphics[width=0.5\linewidth]{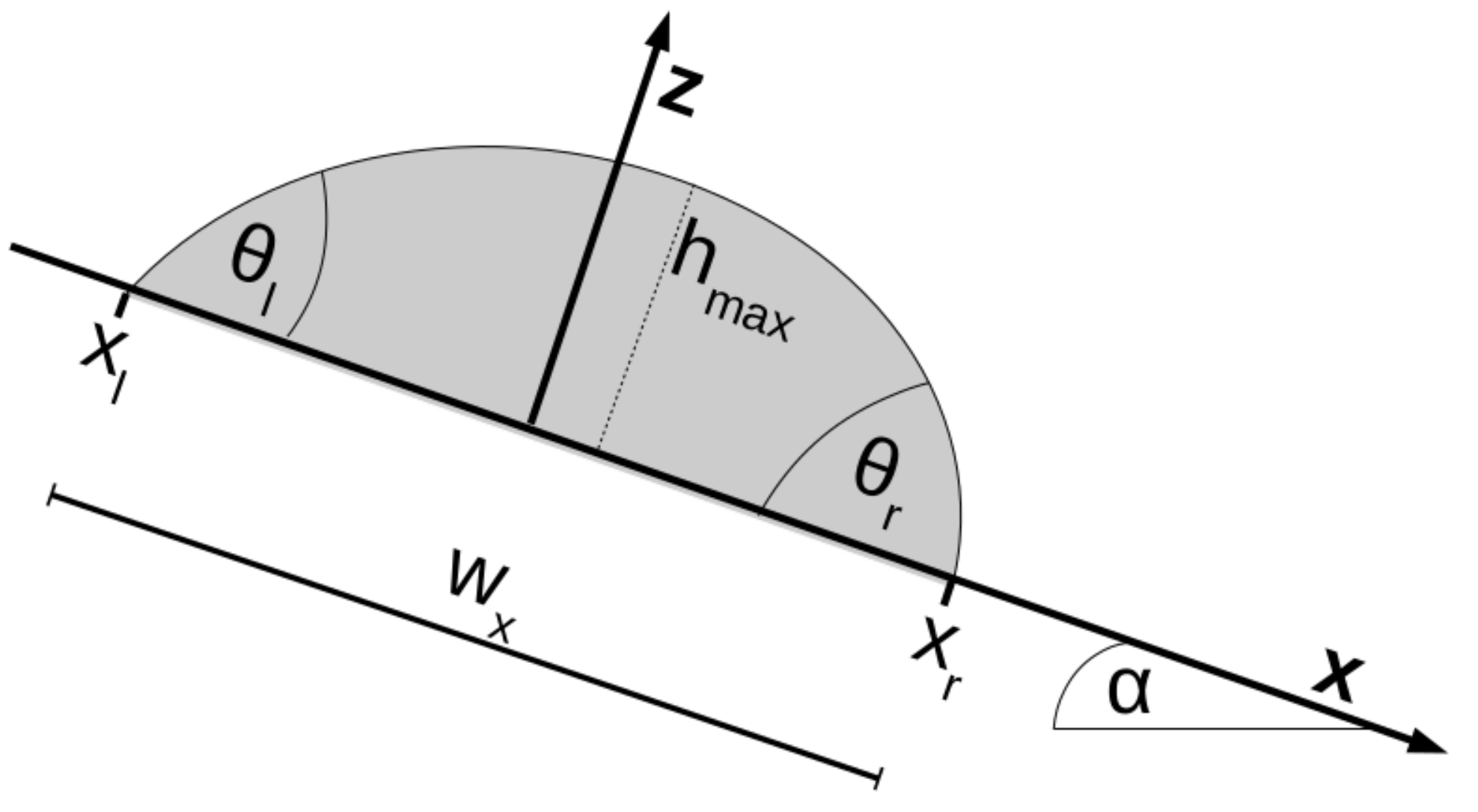}\label{Fig:EsquemaPlanoXZ}
\caption{Sketch of the drop cross section along the symmetry vertical plane on
an incline. }
\label{fig:sketch}
\end{center}
\end{figure}

An aspect that has drawn considerable attention in this context is the shape
that a tilted sessile drop adopts under different plane inclinations, $\alpha$
(see Fig.~\ref{fig:sketch}). This problem can be theoretically tackled by means
of two equivalent approaches, namely, the equilibrium of pressures (by using the
Young--Laplace equation) or the energy minimization method (by considering both
surface and gravitational energies in an Euler--Lagrange framework). The latter
has been considered by using different assumptions, for example, that footprints remains circular $\alpha \neq 0$~\cite{milinazzo_jcis88,brown_jcis80,starov_jpcm09,coninck_pre17}.  
However, there is experimental evidence that the footprints of drops on inclines are not circular. Therefore, some authors have proposed alternative non--circular footprints, such as straight lines and circular endings~\cite{dussan_jfm85}. Although this shape is very simple, and easy to deal with, it does not correspond to observations. A more accurate approximation was developed in~\cite{elsherbini_jcis04b}, where the drop footprint is approximated by two superimposed ellipses sharing a common tangent at the maximum width. However, these proposals are ad--hoc and are not based on physical grounds.

The use of the Surface Evolver simulation tool~\cite{chou_lang12,janardan_csA14,white_lang15,Xu_csA14} is able to yield other non--circular footprints. Alternative approaches numerically obtain the
footprint shapes considering an hybrid diffuse interface with smoothed particle
hydrodynamics model~\cite{das_lang09}, or an equilibrium variation approach
which accounts for a drop's virtual motion on the footprint~\cite{illiev_jcis97}. The experiments in~\cite{berejnov_pre07} for a tilted water drop on a siliconized flat glass slide have shown that there are three transitions of partial depinning, namely: that of the advancing and receding parts of the contact line, and that of the entire contact line leading to the drop’s translational motion. However, as we will see here, the first two stages do not necessarily occur successively as $\alpha$ increases, but an important overlapping of both stages is possible. Chou et al.~\cite{chou_lang12} have compared their theory with the experimental footprints from~\cite{berejnov_pre07}, and the discrepancies were  attributed to the initially non--circular footprint of the experiments. In this work, we will observe that the experiments carried out in~\cite{chou_lang12} correspond only to the initial stages of the complete $\alpha$--cycles reported here. 

\begin{figure}[htb]
\begin{center}
\includegraphics[width=0.5\linewidth]{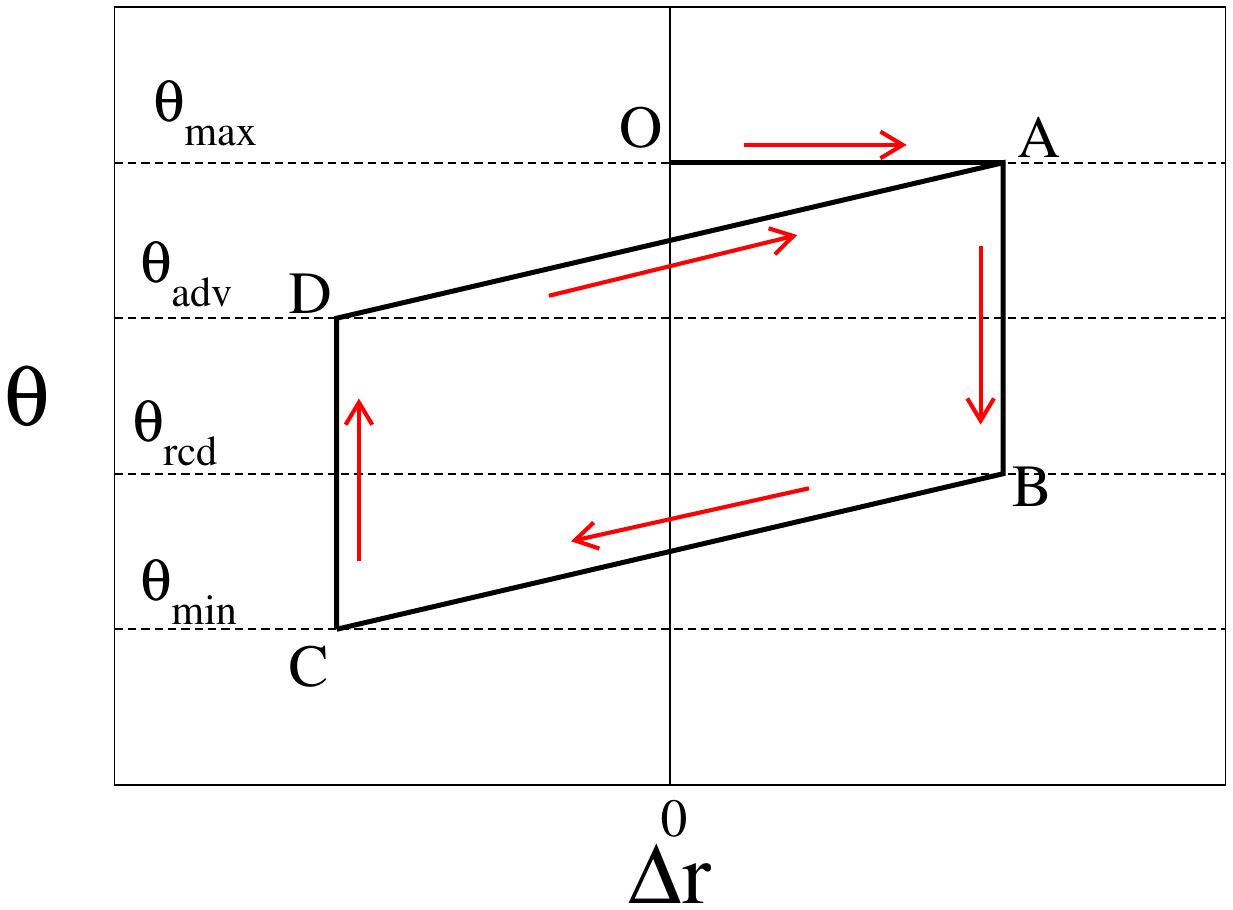}
\caption{Scheme of the contact angle hysteresis cycle. Point O corresponds to
the initial situation where drop volume is $V$ and $\Delta r=0$. The line OA
shows the path to enter the cycle ABCD. Both $V$ and $\theta$ decrease
(increase) from A to C (C to A). Note that there is a pinned contact line ($\Delta
r=const.$) along AB and CD, while there is depinning along BC and DA (see Fig.
2b in~\cite{rava_pre17}).}
\label{fig:sch_hyst}
\end{center}
\end{figure}

The analysis of this deceitfully simple problem is very complex since it
requires a thorough comprehension of the phenomena related to contact angle
hysteresis. In general, this property has been studied by measuring the contact
angle of a sessile drop on a horizontal plane for a given volume, $V$. This is usually 
done by injecting and withdrawing liquid with a needle touching the top of the
drop (see e.g.~\cite{lam_acis02}). A recent study~\cite{rava_pre17} has shown
that, under volume changes, the drop achieves its equilibrium by adjusting the contact angle, 
$\theta$, within a given interval with three different contact line behaviors,
namely: it remains pinned, or it alternatively depins and reaches
equilibrium after having receded or advanced. This cycle is depicted
in Fig.~\ref{fig:sch_hyst}, where $\Delta r$ indicates the displacement of the
equilibrium contact line from the initial situation, marked by the point O where
$\Delta r=0$ (see also Fig.~2b in~\cite{rava_pre17}). The entrance to the cycle
is shown by the path OA along which the drop volume increases while $\theta$ remains constant. 
The right (AB) and bottom (BC) branches result from decreasing
volumes ($\Delta V<0$), while the left (CD) and top (DA) ones correspond to $\Delta V>0$. The three possible behaviors mentioned above are limited by four characteristic contact angles, which define the following three stages: (i) the contact line remains pinned for $\theta_{rcd}<\theta<\theta_{adv}$, (ii) it recedes and reaches equilibrium for $\theta_{min}<\theta<\theta_{rcd}$, and (iii) it advances and reaches equilibrium for $\theta_{adv}<\theta<\theta_{max}$. Note that it is usual to define the hysteresis range only by the interval of case (i), where the contact line is pinned. However, other equilibrium solutions are also found out of this range after some contact line displacements, and they exist within the wider interval
$(\theta_{min},\theta_{max})$. 

In this paper, instead of considering the drop volume, $V$, as the variable
parameter for $\alpha=0$, we vary the inclination angle, $\alpha$, for given
$V$. The goal is to show that a similar hysteresis cycle with the same four
characteristic angles can be found by means of this alternative procedure. Thus, we
wish to confirm that the hysteresis cycle is an intrinsic property of the
liquid--solid interaction, no matter which parameter is varied around it, $V$ or
$\alpha$. In particular, we will start the cycle at $\alpha=0$ and analyze
the drop equilibrium states within the interval $(-\alpha_{max},\alpha_{max})$,
where $\alpha_{max} \lesssim \alpha_{crit}$ (in our case $\alpha_{crit} \approx
26^\circ$). One advantage of the procedure of changing $\alpha$ for fixed $V$ is
that it is not invasive, since the drop is only in contact with the substrate,
and no needle is perturbing its free surface. On the other hand, the axial
symmetry of the procedure with variable $V$ and $\alpha=0^\circ$ is lost, since
the drop footprint is no longer circular as the plane is inclined. This seeming
drawback is here utilized to obtain two hysteresis cycles, one for the left and
another for right contact points of the drop when observed laterally. Note that
these points change their downhill or uphill character as $\alpha$ changes sign.

Moreover, we are also concerned with the modifications of the drop shape as the
inclination varies in the interval $(-\alpha_{max},\alpha_{max})$. In
particular, the experiments show an expected symmetry of the profiles for
positive and negative values of $\alpha$, which can be used an assessment of the
experimental accuracy. We also calculate the theoretical drop shape by
numerically solving the pressure equilibrium equation for the drop thickness,
$h$. The accuracy of the long--wave theory approach is discussed by numerically
solving the simplified equation for $h$. Within this approximation, known as
lubrication theory, we find an analytical solution in the form of a series. In
order to obtain a solution for practical use, we propose a truncated expression, whose 
coefficients can be calculated from a small set of easily measured drop
parameters, namely, the drop extensions along both downhill and transverse
directions, and the downhill contact angle. Using this model, we are able to
predict other measured features of the drop that require more complex
diagnostics, such as the uphill contact angle, the maximum drop thickness, and
the footprint shape. A comparison is made between these predictions and their
measured counterparts. 

The paper is organized as follows. Section~\ref{sec:exp_data} describes the
experimental setup and presents measured data regarding the whole hysteresis
cycle, which can be divided into five main branches. We also analyze the drop
thickness profiles by comparing measurements for positive and negative
$\alpha$'s. Then, we describe the formalism and different approaches to account
for the calculation of the equilibrium drop shape in Section~\ref{sec:drop}. The
numerical solutions for the approaches with and without lubrication
approximation are compared with the experimental data in Section~\ref{sec:num}.
A similar comparison is done in Section~\ref{sec:lubt} for the truncated
solution under the long--wave approximation. Finally, Section~\ref{sec:conclu}
is devoted to a summary and conclusions.

\section{Experiments and description of the hysteresis cycle}
\label{sec:exp_data}

The experiments are performed on a substrate which is partially wetted by our
working fluid, namely a silicone oil (polydimethylsiloxane, PDMS). The substrate
is a glass (microscope slide) coated with a fluorinated solution (EGC-1700 of
3M) by controlling both the dip coating velocity ($\approx 0.1$~cm/min) and the
drying time of the solvent solution ($t \approx 30$~min) in an oven at
temperature $T\approx 40^\circ C$. Under this protocol we ensure that the
PDMS partially wets the substrate in a repeatable way, so that a drop of given
volume placed on a horizontal substrate always reaches the same contact angle
(within the experimental error, $\pm 0.5^\circ$). Surface tension, $\gamma$, and
density, $\rho$, of PDMS are measured with a Kr\"uss K11 tensiometer, while its
viscosity, $\mu $, is determined with a Haake VT550 rotating viscometer. The
values of these parameters are: $\gamma = 21$~dyn/cm, $\rho = 0.97$~g/cm$^{3}$,
and $\mu = 21.7$~Poise at temperature $T=20^\circ C$.

The same kind of substrate and PDMS was previously used in~\cite{rava_pre17},
where the the static contact angle, $\theta$, of a single sessile drop with
circular footprint on a horizontal substrate was measured. In that work, we were
able to study the hysteresis cycle of $\theta$ by injecting and withdrawing
fluid with a needle in contact with the top of the drop. This was done by
measuring $\theta$ (as well as the thickness profile, $h(x)$) with a Rame--Hart
Model 250 goniometer. In particular, we found $\theta_{max}=55^\circ$ and
$\theta_{min}=40^\circ$, while the contact angles at which the contact line
displaces to achieve equilibrium, that is the advancing and receding angles,
were $\theta_{adv}=52^\circ$ and $\theta_{rcd}=46^\circ$, respectively. 

\begin{figure}[htb]
\begin{center}
\includegraphics[width=0.55\linewidth]{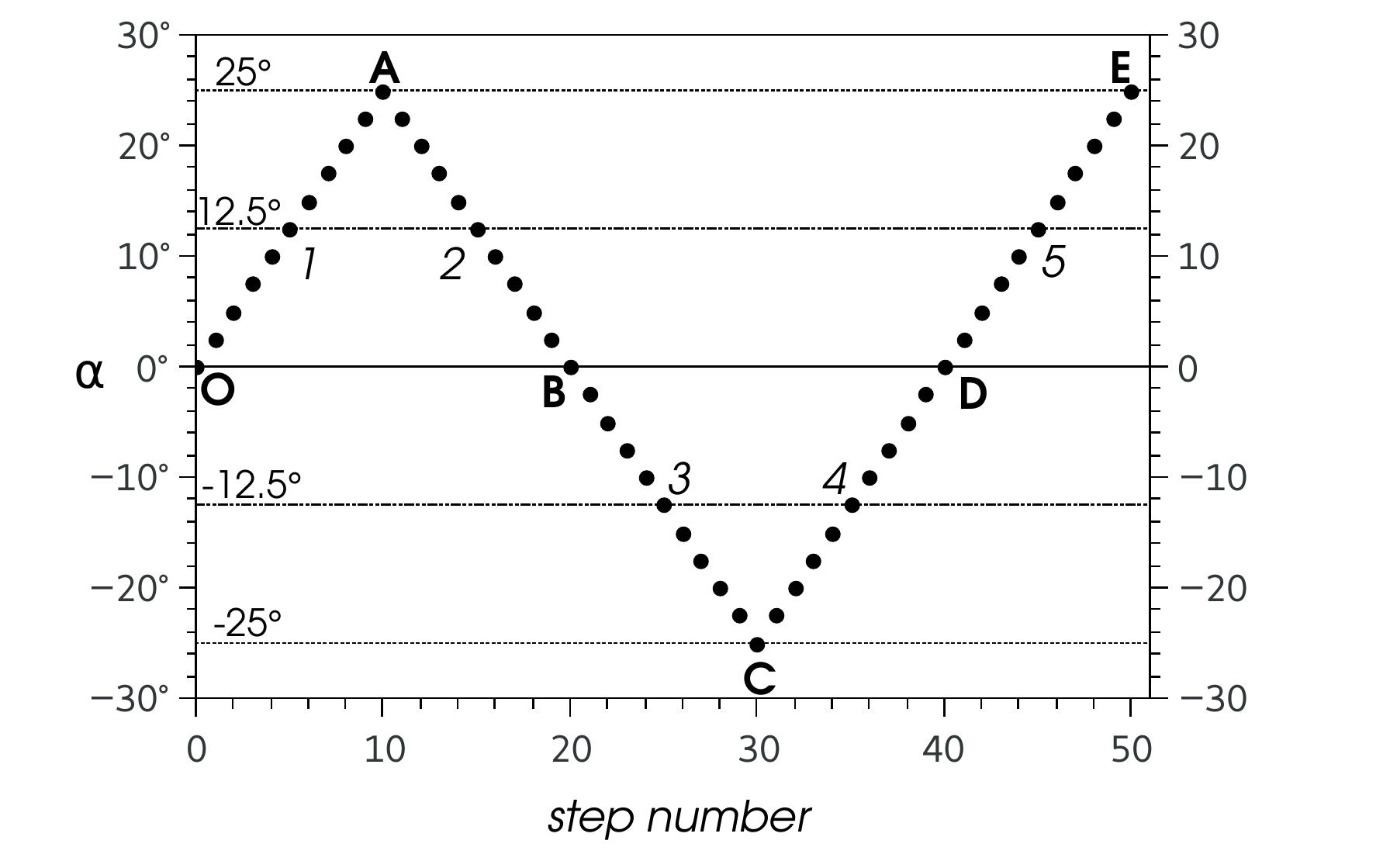}
\caption{Inclination angle, $\alpha$, as a function of the number of steps for $\alpha_{max}=25^\circ$. The complete cycle is: $0^\circ \rightarrow 25^\circ \rightarrow 0^\circ
\rightarrow -25^\circ \rightarrow 0^\circ \rightarrow 25^\circ$. Each step corresponds to $\Delta
\alpha=2.5^\circ$. The numbers and letters indicate the branches and their
extreme points, respectively.}
\label{fig:alphas}
\end{center}
\end{figure}

Unlike the experiments in~\cite{rava_pre17}, the goniometer is now mounted on a
tilted base with a variable inclination angle, $\alpha$. Since the goal here is
to analyze the deformations of the drop shape due to changes in $\alpha$, the
substrate is inclined in successive steps of $\Delta \alpha=2.5^\circ$, and the
cycle $0^\circ \rightarrow \alpha_{max} \rightarrow 0^\circ \rightarrow
-\alpha_{max} \rightarrow 0^\circ \rightarrow \alpha_{max}$, with
$\alpha_{max}<\alpha_{crit}\approx 26^\circ$ is considered (see 
Fig.~\ref{fig:alphas}) where the branches are numbered from $1$ to $5$, the
starting point is denoted by O, and the extreme points by A, B, C, D and E. We
measure the positions of the contact line points on the {\emph right}, $x_r$,
and {\emph left}, $x_l$, of the drop profile at the $xz$--plane (see
Fig.~\ref{fig:sketch}) as well as the corresponding contact angles, $\theta_r$ and $\theta_l$. 
Note that $x_r$ corresponds to the downhill (uphill) point for $\alpha>0$ ($\alpha<0$), and vice
versa for $x_l$. 

\begin{figure}[htb]
\centering
\subfigure[]{\includegraphics[width=0.3\textwidth]{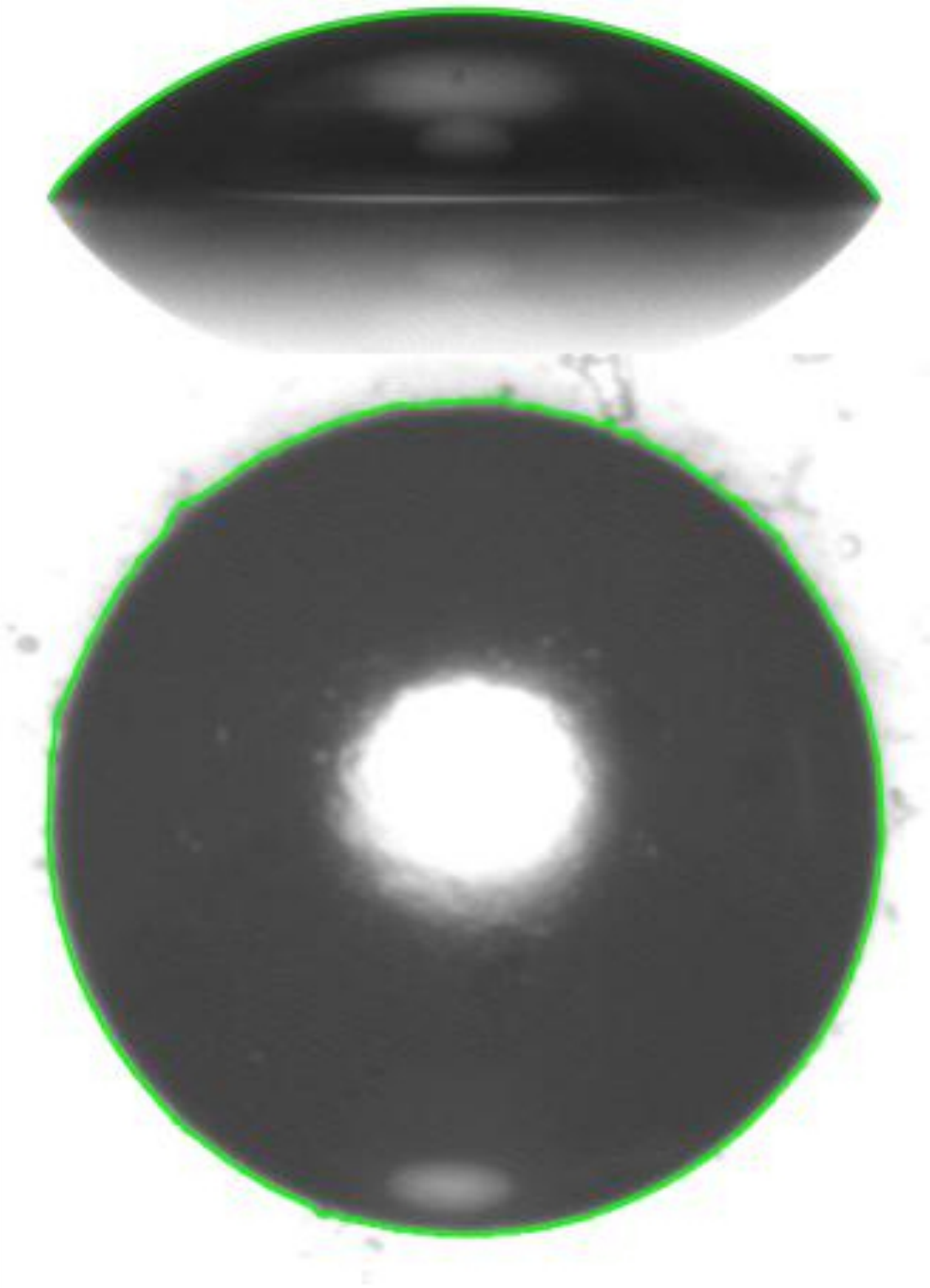}}
\subfigure[]{\includegraphics[width=0.3\textwidth]{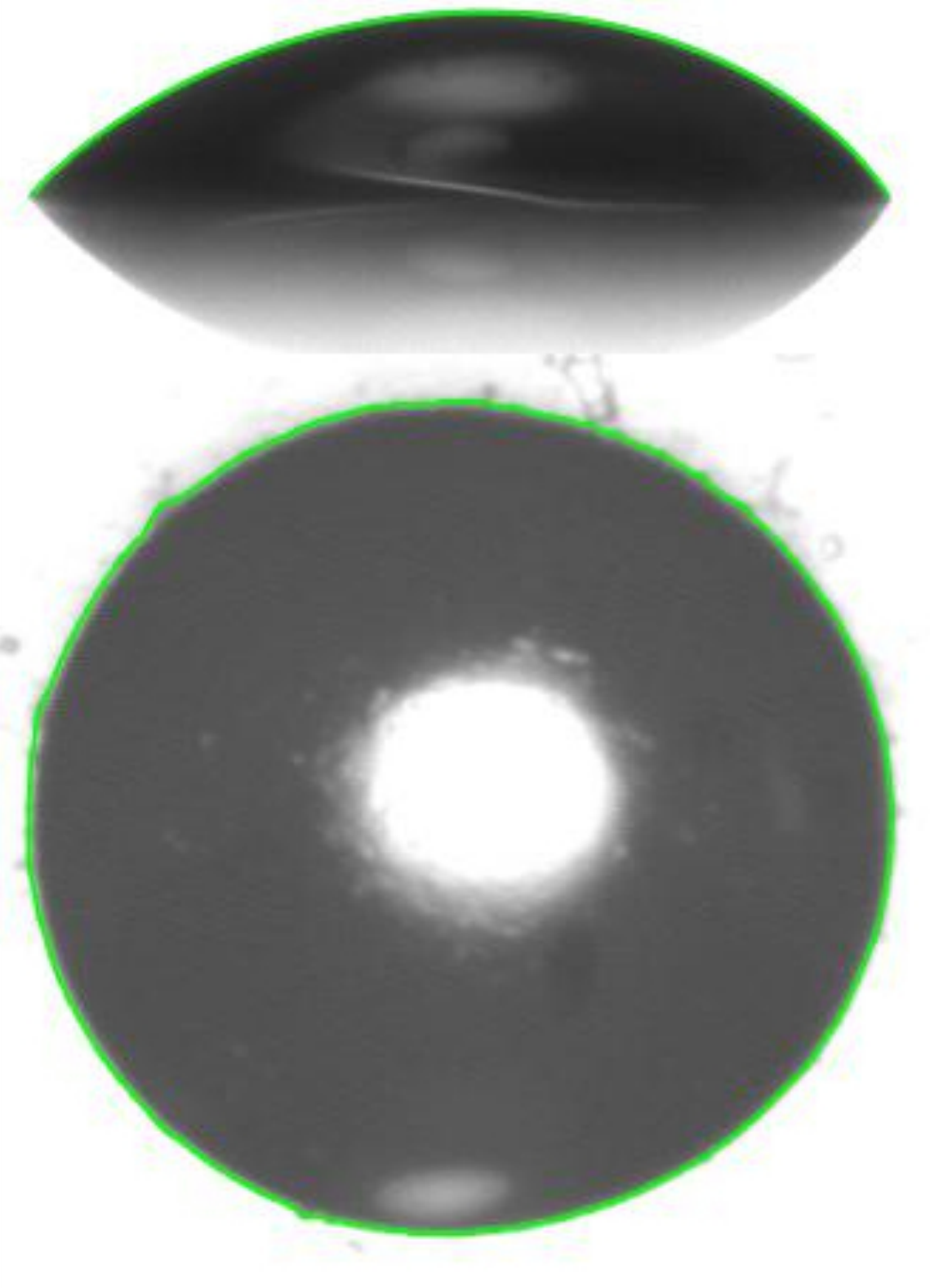}}
\subfigure[]{\includegraphics[width=0.3\textwidth]{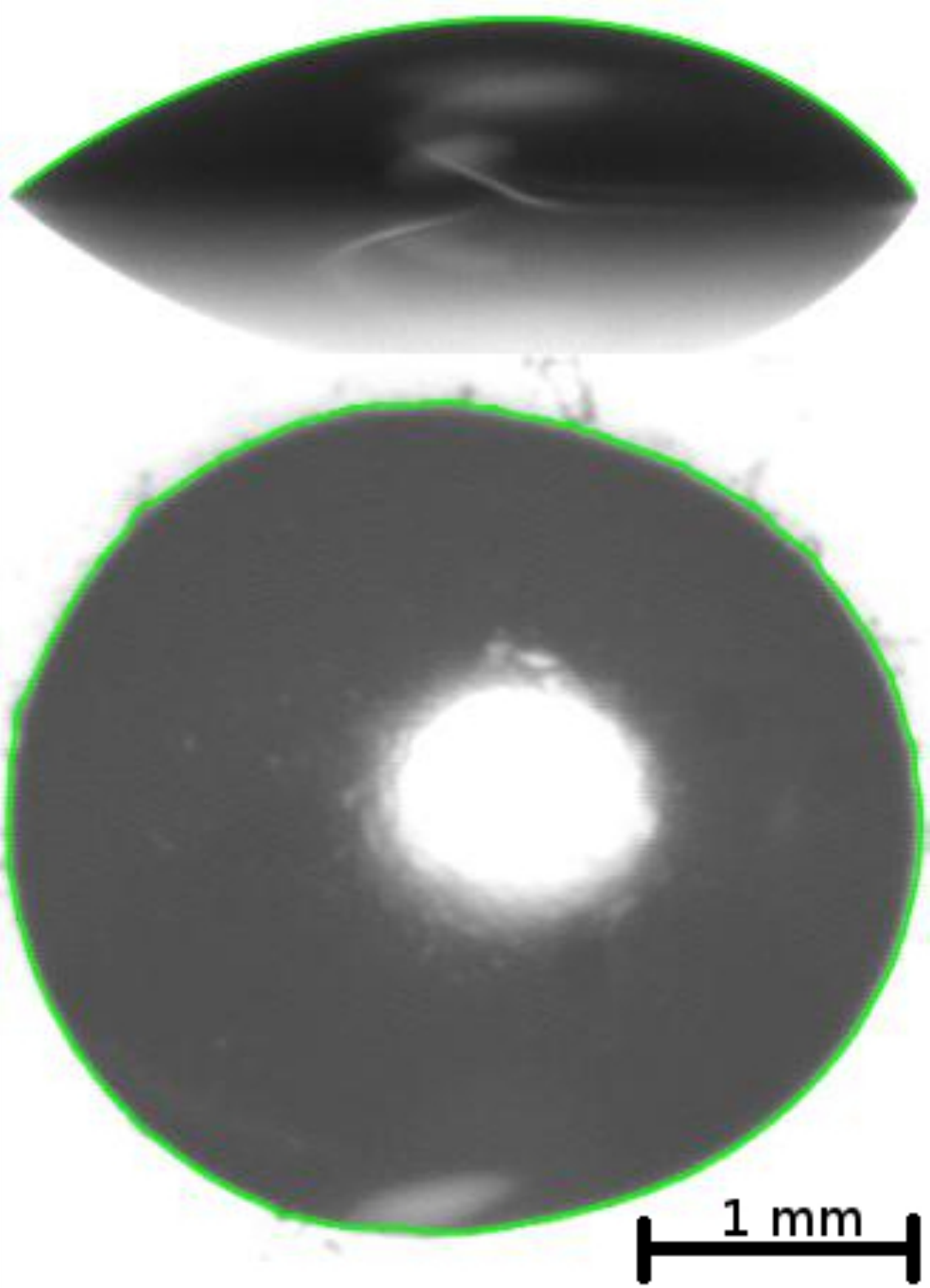}}
\caption{Images as obtained from the goniometer of the side (top line) and top
(bottom line) views of the drop for three inclination angles, $\alpha$. (a) $\alpha=0^\circ$, (b) $\alpha=12.5^\circ$, and (c) $\alpha=25^\circ$. The green lines are the contours extracted from the image analysis. The bar in (c) gives the scale of all pictures.}
\label{fig:images}
\end{figure}

The thickness profiles and footprint shapes at every point of the cycle are
obtained from images such as those shown in Fig.~\ref{fig:images}. The top
(bottom) line shows the side (top) views of the drop for
$\alpha=0^\circ$,$12.5^\circ$, and $25^\circ$. The reflected image of the drop
on the substrate, as seen at the top line of the pictures, is used to determine
the substrate position as well as the segment connecting the side vertices (its
length yields the drop extension, $w_x$). The thick green lines in the top line
pictures indicate the thickness profiles, while those in the top views (bottom
line) correspond to the footprint shapes.

The initial contact angle at point O, $\theta_0$, depends on how the drop is set
up on the substrate. For this task, we use an automatic dispenser syringe; thus,
$\theta_0$ may result in the range $\theta_{adv}<\theta_0<\theta_{max}$ since
the drop reaches equilibrium by spreading after being deposited with the needle.
Usually, it results $\theta_0=\theta_{max}$, but it can be reduced a bit if some
drop fluid is withdrawn with the needle. In the following, we will analyze the
effects of both $\alpha_{max}$ and $\theta_0$ on the hysteresis cycle.

\begin{figure}[htb]
\centering
\subfigure[]{\includegraphics[width=0.49\textwidth]{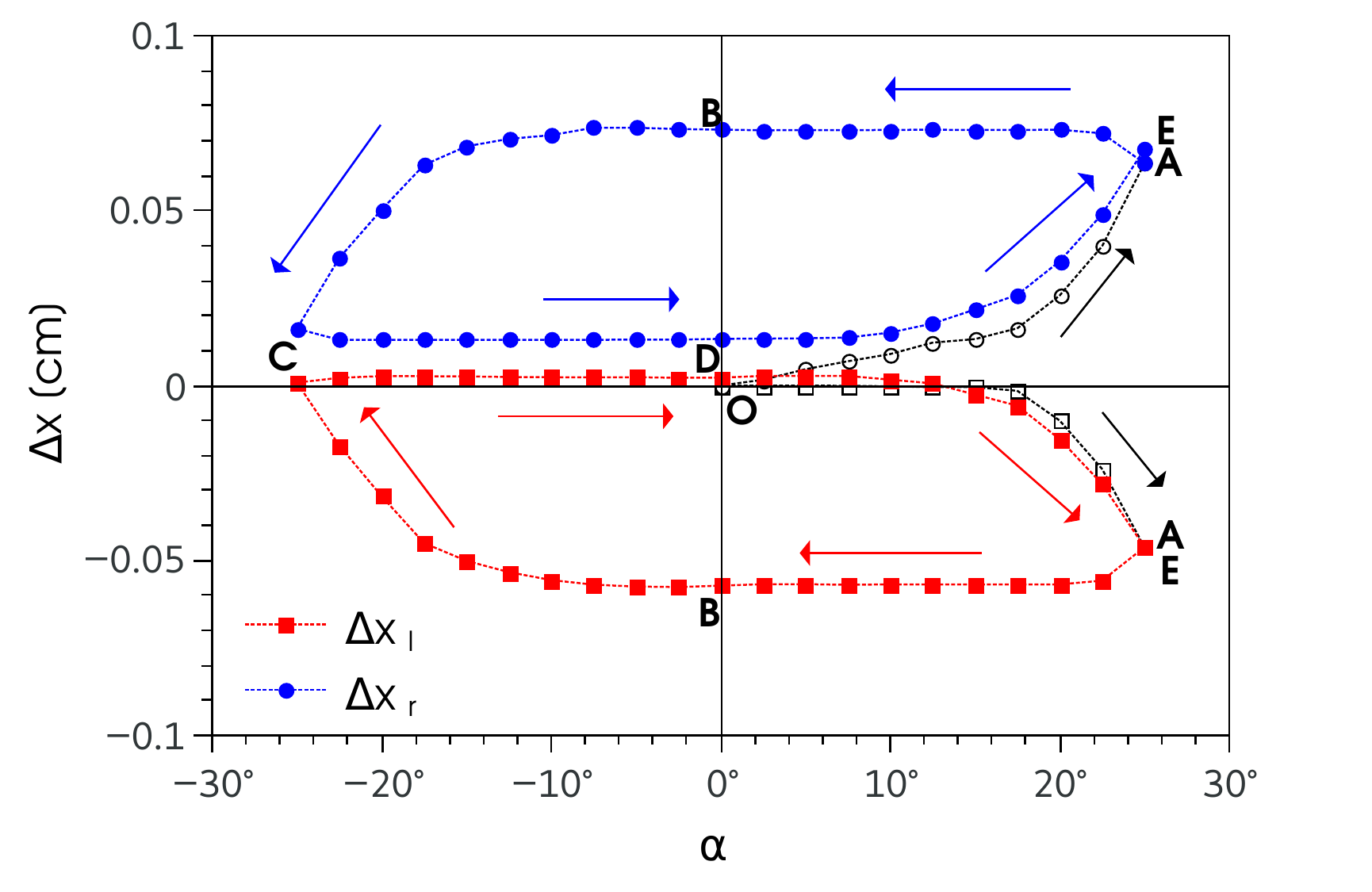}}
\subfigure[]{\includegraphics[width=0.49\textwidth]{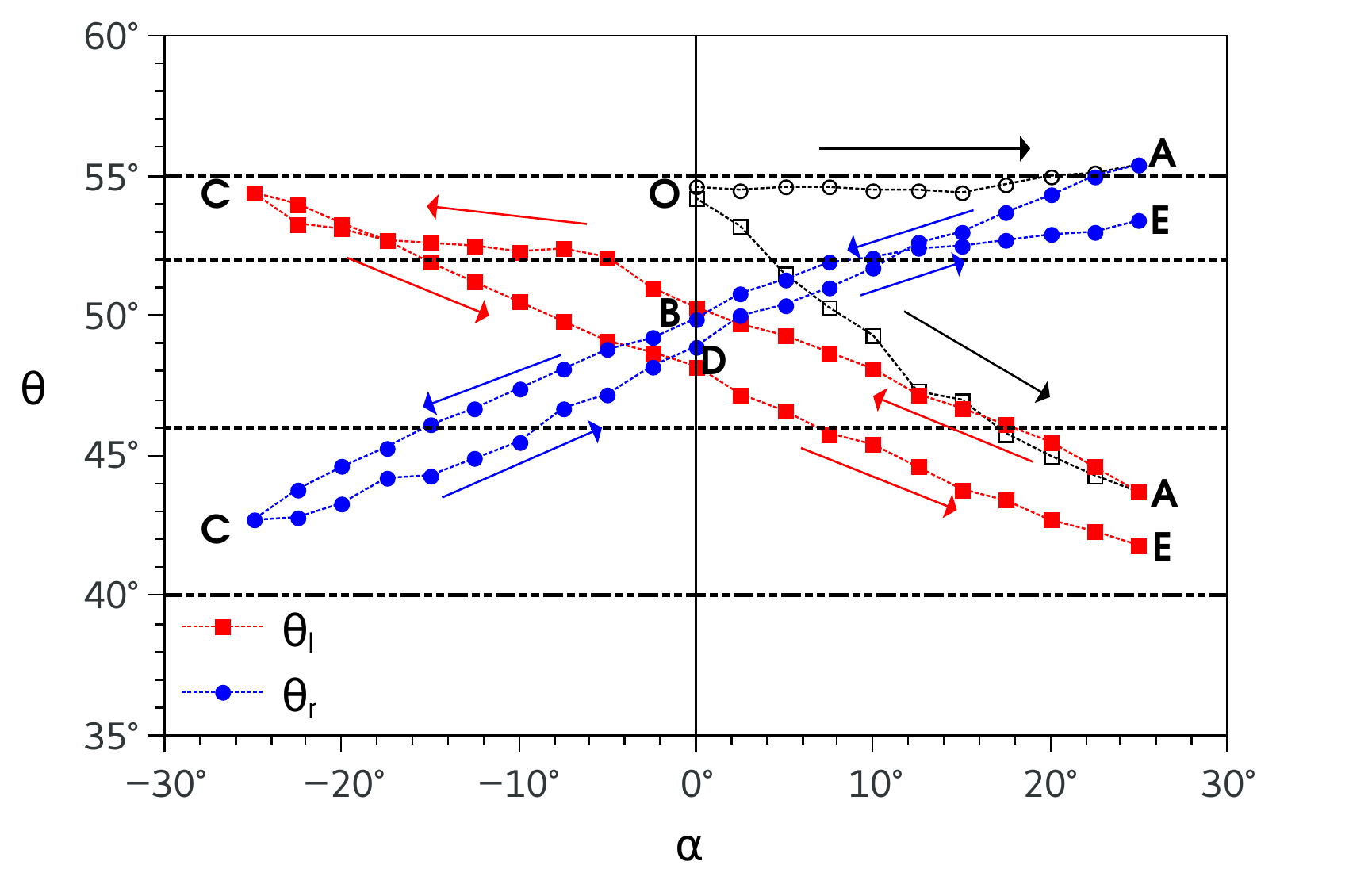}}
\caption{Maximum plane inclination $\alpha_{max}=25^\circ$ and initial contact
angle $\theta_0=55^\circ$: (a) Contact lines displacements $\Delta x_l$ and $\Delta x_r$ (with respect to their positions at $\alpha=0^\circ $) versus $\alpha$. (b) Contact angles $\theta_l$ and $\theta_r$
versus $\alpha$. The horizontal dotted (dot--dashed) lines in (b) correspond to  $\theta_{rcd}=46^\circ$ and $\theta_{adv}=52^\circ$ ($\theta_{min}=40^\circ$ and $\theta_{max}=55^\circ$). The drop volume is $V= 0.657 a^3= 2.17\,mm^3$.} 
\label{fig:a25_thei55}
\end{figure}

We start our study with $\alpha_{max}=25^\circ$ and $\theta_0=\theta_{max}=55^\circ$. The measured displacements $\Delta x_l$ and $\Delta x_r$ (with respect to their positions for $\alpha=0^\circ$) versus $\alpha$ are shown in Fig.~\ref{fig:a25_thei55}a by squares and circles, respectively. The accuracy of these displacements is of $\pm 1$ pixel which,  according to our optical magnification, corresponds to $\pm 11\mu m$. The main $\alpha$--sequence represented in Fig.~\ref{fig:alphas} yields the cycles ABCDE for both $\Delta x_l$ (full red squares) and $\Delta x_r$ (full blue circles) as shown in Fig.~\ref{fig:a25_thei55}a. The entry paths, OA, to the cycles are shown by hollow black squares and circles, respectively. Analogous considerations apply for $\theta_l$ and $\theta_r$ in Fig.~\ref{fig:a25_thei55}b. Note that only the path OA is similar to the experiments reported in Section 3.B of~\cite{chou_lang12}, where no detailed description of the footprints shape was provided.

\begin{figure}[htb]
\centering
\includegraphics[width=0.7\textwidth]{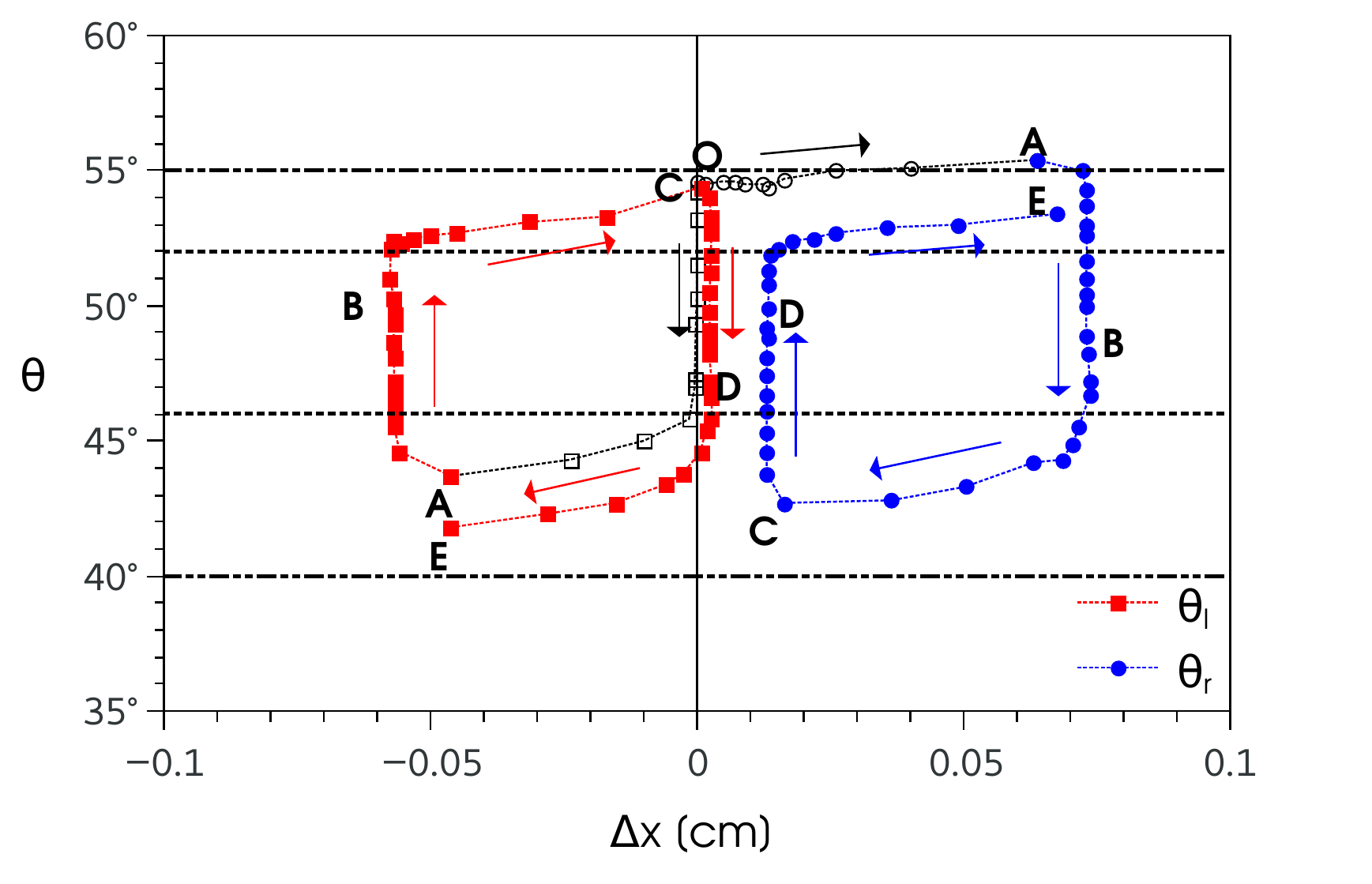}
\caption{Contact angles, $\theta_l$ and $\theta_r$, versus contact line displacements, $\Delta x_l$ and $\Delta x_r$, for $\alpha_{max}=25^\circ$ and $\theta_0=55^\circ$ for the drop in Fig.~\ref{fig:a25_thei55}. The horizontal dotted (dot--dashed) lines correspond to  $\theta_{rcd}=46^\circ$ and $\theta_{adv}=52^\circ$ ($\theta_{min}=40^\circ$ and $\theta_{max}=55^\circ$).}
\label{fig:theta_dx_25_55}
\end{figure}

Consequently, by considering $\alpha$ as a varying parameter, we are able to construct the contact angle hysteresis cycles, as shown in Fig.~\ref{fig:theta_dx_25_55}. Clearly, this cycle has properties similar to those described for the one depicted in Fig.~\ref{fig:sch_hyst}, which uses the drop volume $V$ as a varying parameter. Moreover, the values of the angles limiting the pinning and depinning regions are coincident with those in~\cite{rava_pre17}, where the same kind of substrate and fluid were used.

\begin{figure}[htb]
\centering
\subfigure[]{\includegraphics[width=0.49\textwidth]{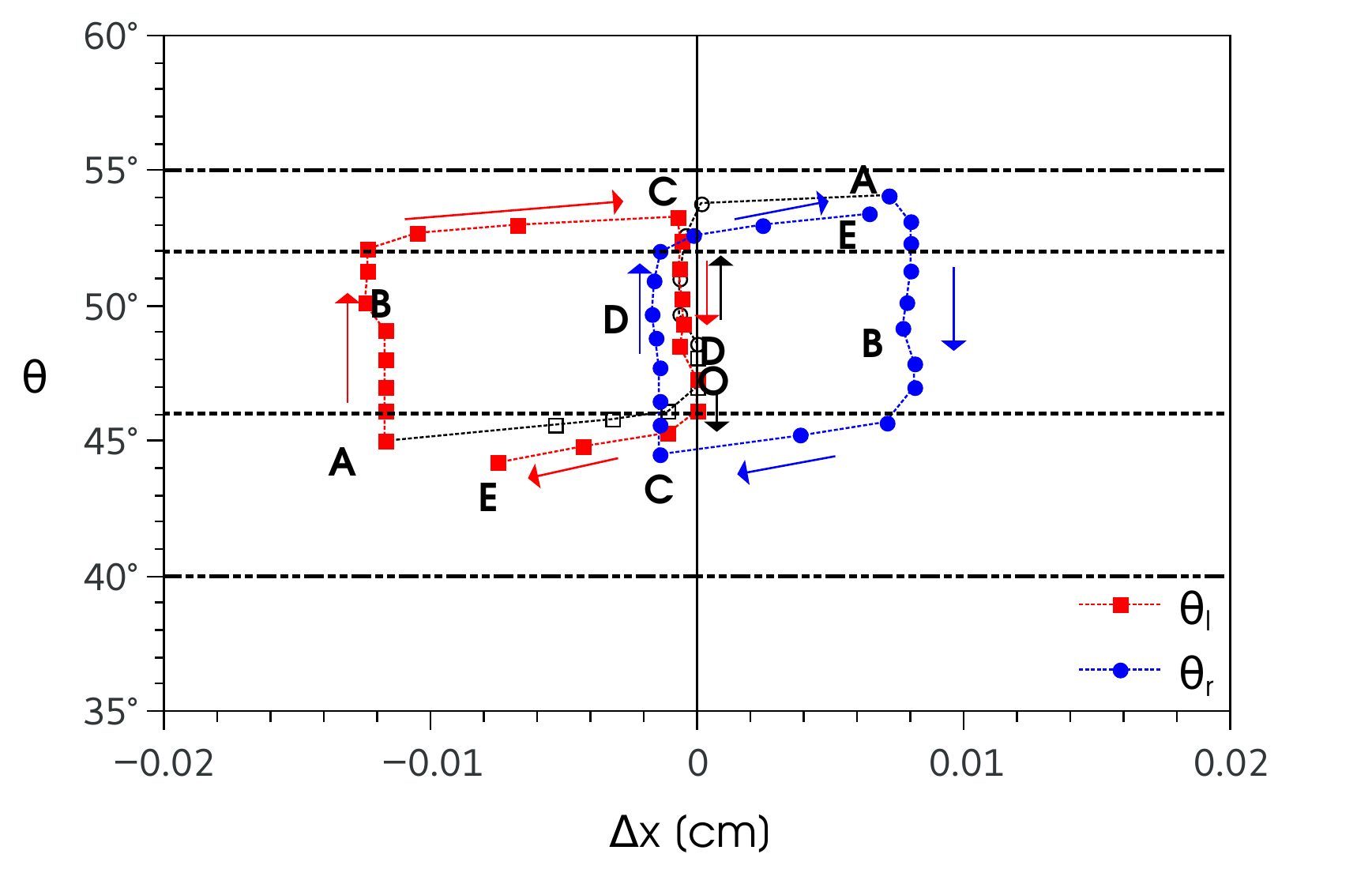}}
\subfigure[]{\includegraphics[width=0.49\textwidth]{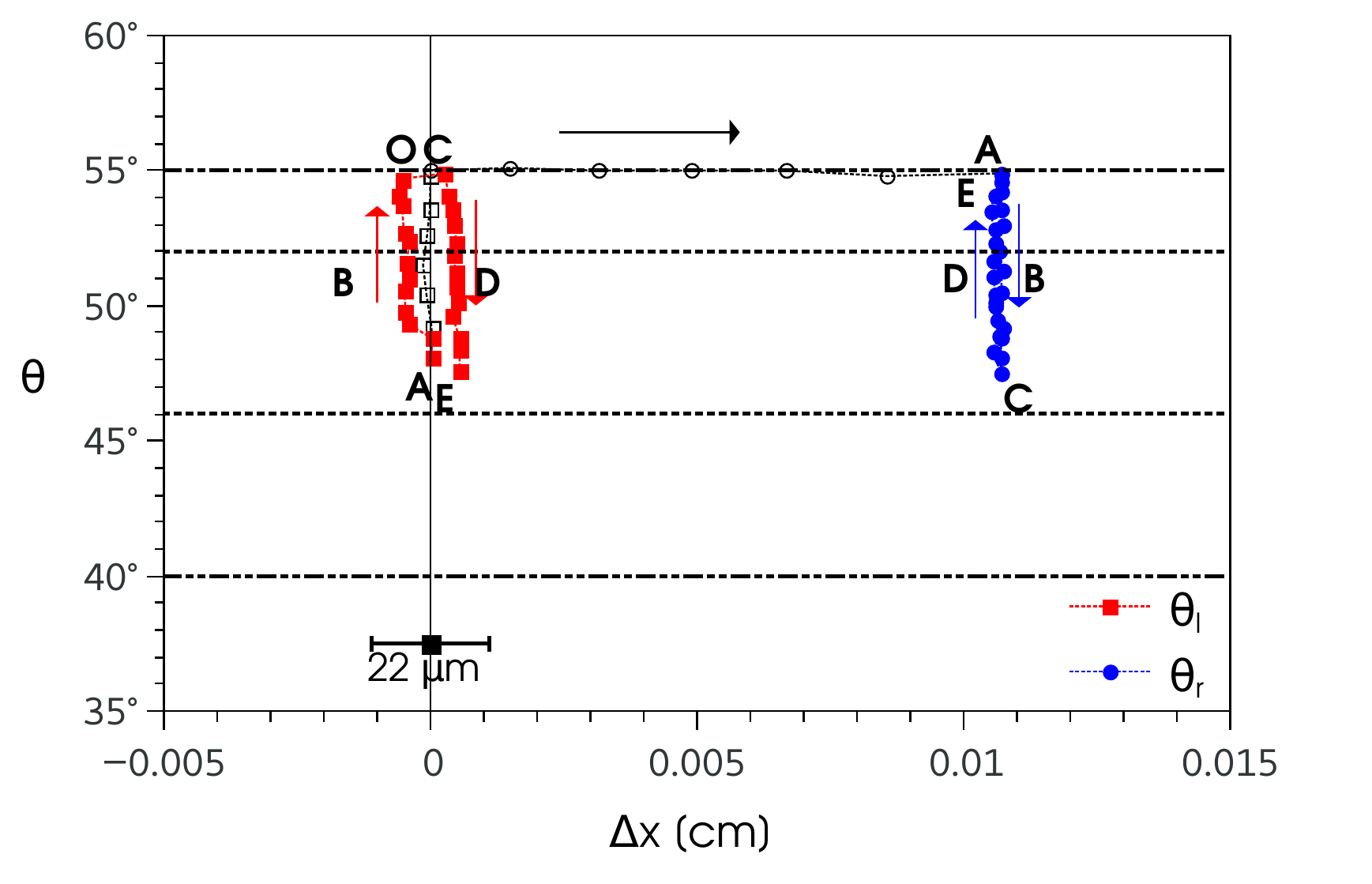}}
\caption{Contact angles, $\theta_l$ and $\theta_r$, versus contact line displacements, $\Delta x_l$ and $\Delta x_r$, for: (a) $\alpha_{max}=25^\circ$ and $\theta_0=48^\circ$ ($V= 0.474 a^3= 1.57\,mm^3$). (b) $\alpha_{max}=15^\circ$ and $\theta_0=55^\circ$ ($V= 0.519 a^3= 1.72\,mm^3$). The horizontal segment in (b) corresponds to the error bar for $\Delta x$ .}
\label{fig:theta_dx}
\end{figure}

Equivalent cycles are also obtained for smaller $\theta_0=48^\circ$ and same $\alpha_{max}$ than in Fig.~\ref{fig:theta_dx_25_55}, i.e. for $\theta_0<\theta_{adv}$ (see Fig.~\ref{fig:theta_dx}a). The $\Delta x$--range of the cycles decreases significantly (from $\approx 0.06$~cm to $\approx 0.01$~cm), but the $\theta$--range is only slightly smaller and scarcely out of the pinning interval $(\theta_{rcd},\theta_{adv})$. This is because the equilibrium drop states along the OA branch can be reached just by increasing (decreasing) $\theta_r$ ($\theta_l$) as $\alpha$ increases without need to displace the contact line. These behaviors are similar to those shown in Section 3.C of~\cite{chou_lang12}. In fact, the displacements occur after $\theta_r$ ($\theta_l$) has approached to $\theta_{adv}$ ($\theta_{rcd}$), i.e. when the contact angles reach the boundaries of the pinning interval $(\theta_{rcd},\theta_{adv})$. This description, based on $\theta$ versus $\Delta x$, is quite general and has the advantage that it avoids the dependence on $\alpha$, in contrast to the experimental results analysis performed in~\cite{janardan_csA14}. Note that the value of $\alpha$ at which $\theta_r$ reaches $\theta_{adv}$ (namely, $\alpha_m$ in~\cite{janardan_csA14}) is not independent of $\theta_0$. Moreover, the numerical results in~\cite{white_lang15} show that $\alpha_1$ ($=\alpha_m$ in~ \cite{janardan_csA14}) depends not only on $\theta_0$ but also on the drop volume, $V$. On the other hand, the results in Figs.~\ref{fig:theta_dx_25_55} and \ref{fig:theta_dx} show that the transitions regions in the hysteresis cycle are intrinsic properties of the system, and therefore independent of both $\theta_0$ and $V$.

In Fig.~\ref{fig:theta_dx}b, we keep the same initial condition as in Fig.~\ref{fig:theta_dx_25_55} and reduce the value of $\alpha_{max}$ to $15^\circ$. We observe that $\Delta x_r>0$ only along the initial branch OA, while the cycles show variations of both $\theta_r$ and $\theta_l$ with $\Delta x_r=\Delta x_l=0$, i.e. with pinned contact lines. The reason for this behavior is that for such small $\alpha$, the equilibrium states can be reached for $\theta_r$ and $\theta_l$ greater than $\theta_{rcd}$, without need to displace the contact lines.   

\subsection{Branch 1: From $\alpha=0$ to $\alpha=25^\circ$ (O $\rightarrow$ A
with $\alpha_{max}=25^\circ$ and $\theta_0=55^\circ$)}
\label{sec:branch1}

As $\alpha$ increases from $0$ to $25^\circ$ we observe that the downhill
contact point, $x_r$, displaces downwards a bit more than the uphill one, $x_l$ (see
Fig.~\ref{fig:a25_thei55}a, O $\rightarrow$ A for both $\Delta x$'s).
Concomitantly, the downhill contact angle approaches $\theta_{max}$ and remains
practically constant until point A (see Fig.~\ref{fig:a25_thei55}b), while the
uphill contact angle, $\theta_l$, significantly diminishes, surpassing
$\theta_{rcd}$ and approaching $\theta_{min}$ (see Fig.~\ref{fig:a25_thei55}b).
These facts indicate that $\alpha=25^\circ$ is very close to the limiting angle,
$\alpha_{crit}$, beyond which the drop would not remain at rest and would slide
down the plane instead. 

\begin{figure}[htb]
\centering
\subfigure[]{\includegraphics[width=0.55\textwidth]{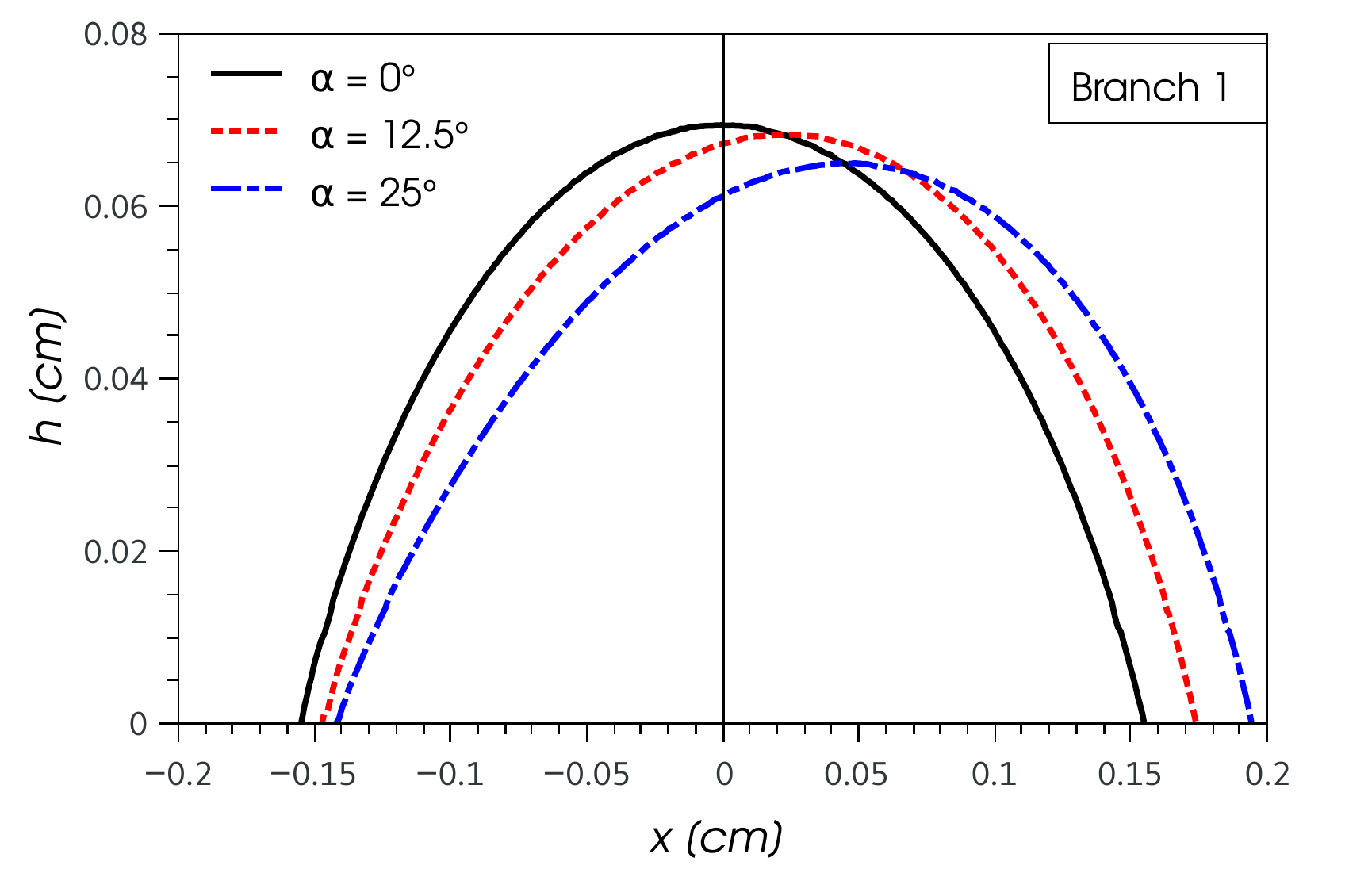}}
\subfigure[]{\includegraphics[width=0.42\textwidth]{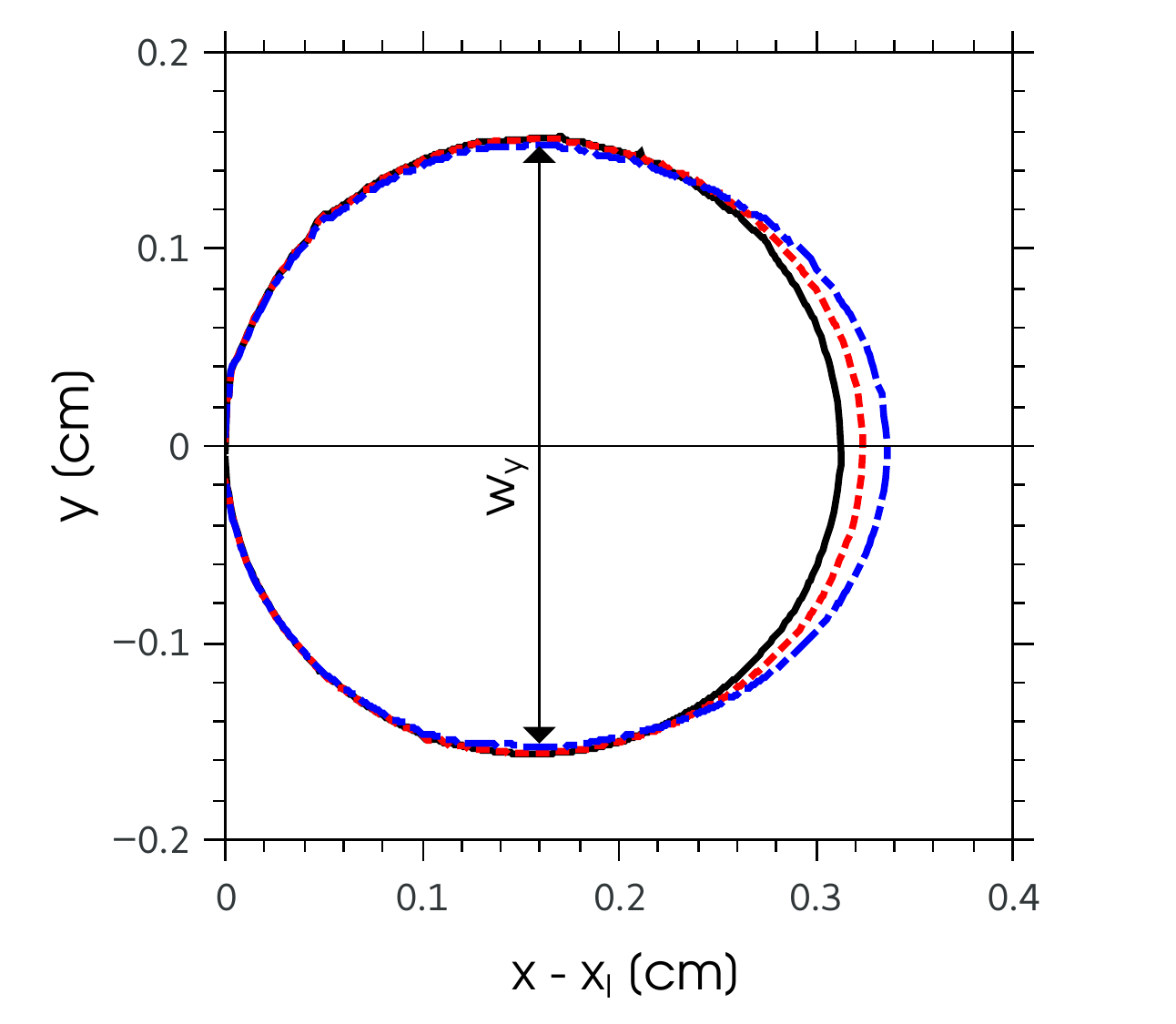}}
\caption{Branch $1$: (a) Thickness profiles at $\alpha=0^{\circ}$ (solid lines),
$12.5^{\circ}$ (dashed lines), and $25^{\circ}$ (dot--dashed line). (b)
Corresponding footprints shifted so that the most left points are coincident. The drop volume is $V=0.876 a^3=2.9\,mm^3$}
\label{fig:hxFoot}
\end{figure}

The experimental thickness profiles shown in Fig.~\ref{fig:hxFoot}a indicate
that the drop width, $w_x$, increases while the maximum thickness decreases as
the plane becomes more inclined. The footprints at points O, A and an
intermediate point $\alpha=12.5^\circ$ are displaced to coincide at their most
left point and compared in Fig.~\ref{fig:hxFoot}b. Note that the footprints are
distorted from a circular shape only in the frontal part, i.e. for $x > w_x/2$,
where they become more elongated. Interestingly, the maximum transversal width
of the footprints, $w_y$, remains constant for all $\alpha$'s.

\begin{figure}[htb]
\centering
\subfigure[]{\includegraphics[width=0.49\textwidth]{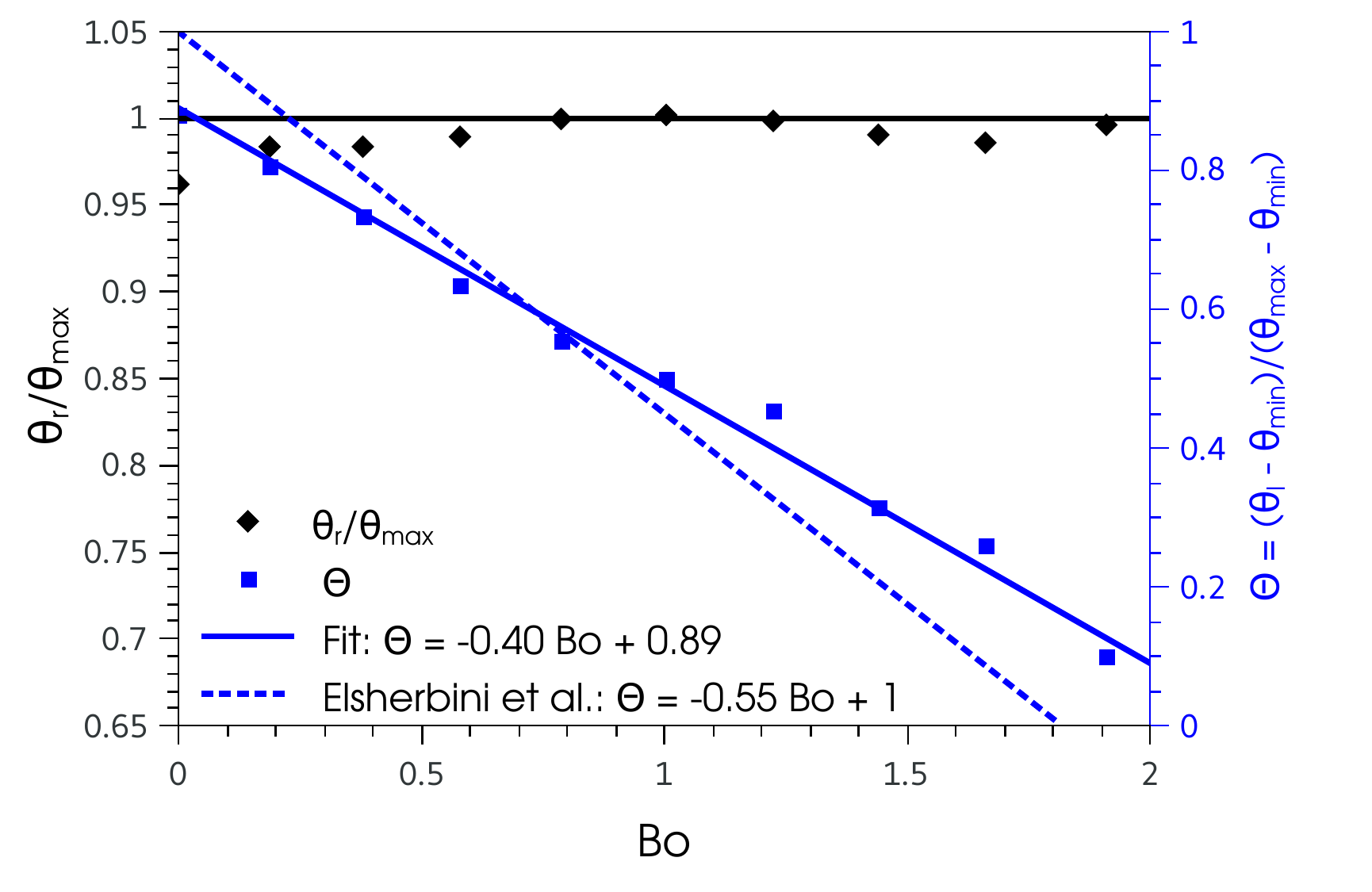}}
\subfigure[]{\includegraphics[width=0.49\textwidth]{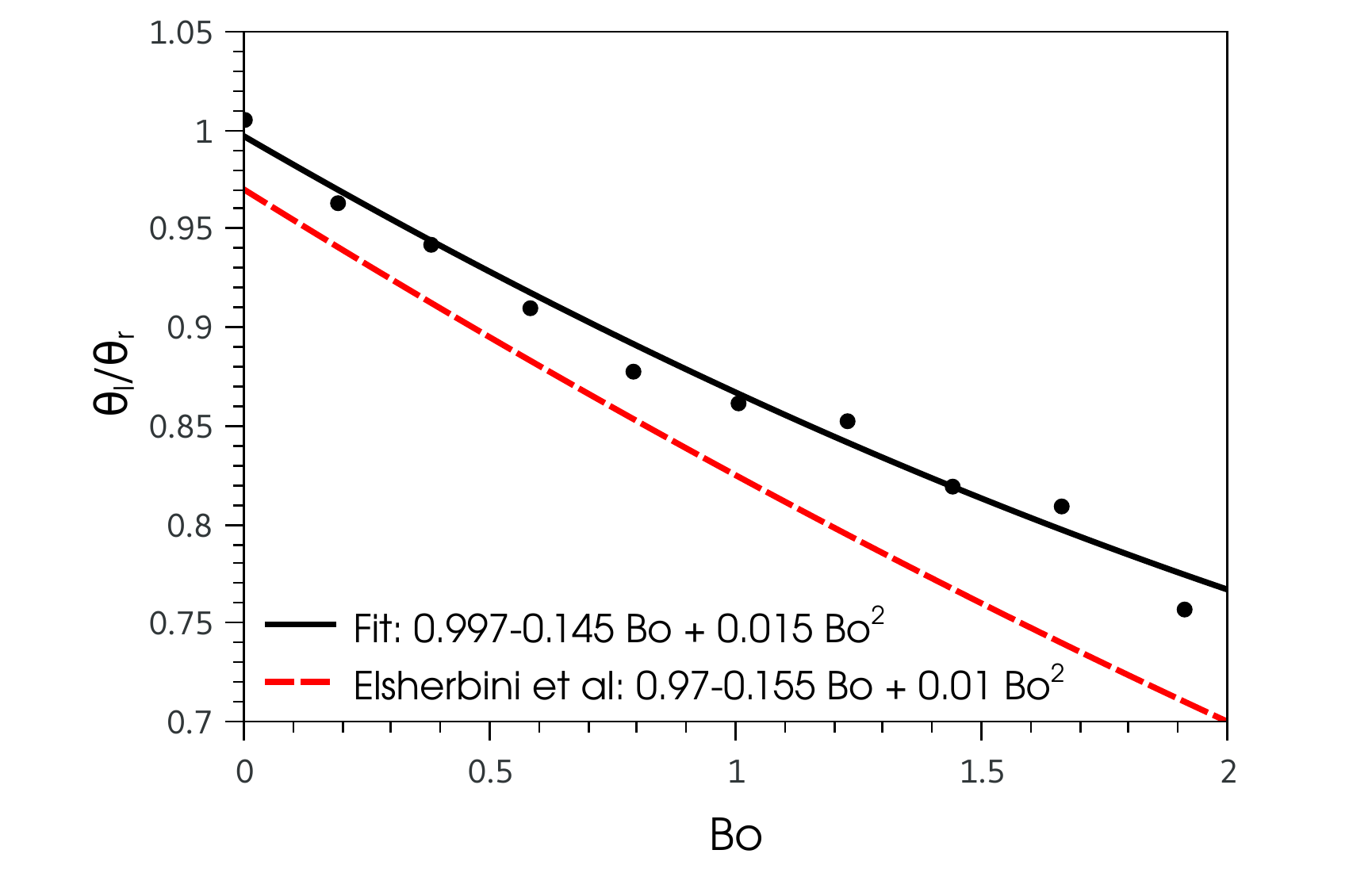}}
\caption{Contact angles as functions of Bond number, $Bo$, along Branch 1: (a)
$\theta_r/\theta_{max}$ (diamonds) and the relative deviation of $\theta_l$:
$\Theta=(\theta_l-\theta_{min}/(\theta_{max}-\theta_{min})$ (squares). The thick
solid line is the best quadratic fit of the squares.(b) Experimental data and
best quadratic fit for ratio of both contact angles $\theta_l/\theta_r$. The
dashed line corresponds to~\cite{elsherbini_jcis04}.}
\label{fig:ratio_Theta}
\end{figure}

Our experimental data for this branch can be compared with those reported in the
literature. In order to do that, we parameterize the effects of the component of
gravity along the plane by defining the Bond number as
\begin{equation}
 Bo = \frac{w_x^2}{a^2} \sin \alpha,
 \label{eq:Bond}
\end{equation}
where $a=\sqrt{\gamma/(\rho g)}$ is the capillary distance. Note that $w_x$ is
also a function of $\alpha$ (see Fig.~\ref{fig:hxFoot}b). For instance, Fig.~\ref{fig:ratio_Theta}a shows that the ratio $\theta_r/\theta_{max}$ (diamonds) remains very close to unity as the plane is
inclined. This result is in agreement with previously reported
experimental~\cite{elsherbini_jcis04,berejnov_pre07} and
numerical~\cite{chou_lang12} results. Analogously to what was done
in~\cite{elsherbini_jcis04}, we plot $\theta_l$ versus $Bo$ by considering its
relative deviation from $\theta_{min}$ as
$\Theta=(\theta_l-\theta_{min})/(\theta_{max}-\theta_{min})$. Similarly to what
was reported in~\cite{elsherbini_jcis04}, we also find a linear dependence for
this relationship. Slopes differences are a consequence of dealing with
different liquid--substrate combinations. In Fig.~\ref{fig:ratio_Theta}b we show the ratio $\theta_l/\theta_r$ versus $Bo$ and obtain a quadratic fitting function (thick solid line). Interestingly, this
curve is similar to that reported in~\cite{elsherbini_jcis04} for other
combinations of liquids and surfaces (see dashed line and formulae in the
figure).

\subsection{Branches 2, 3 and 4}
\label{sec:branch2}
\begin{figure}[htb]
\centering
\subfigure[]{\includegraphics[width=0.55\textwidth]{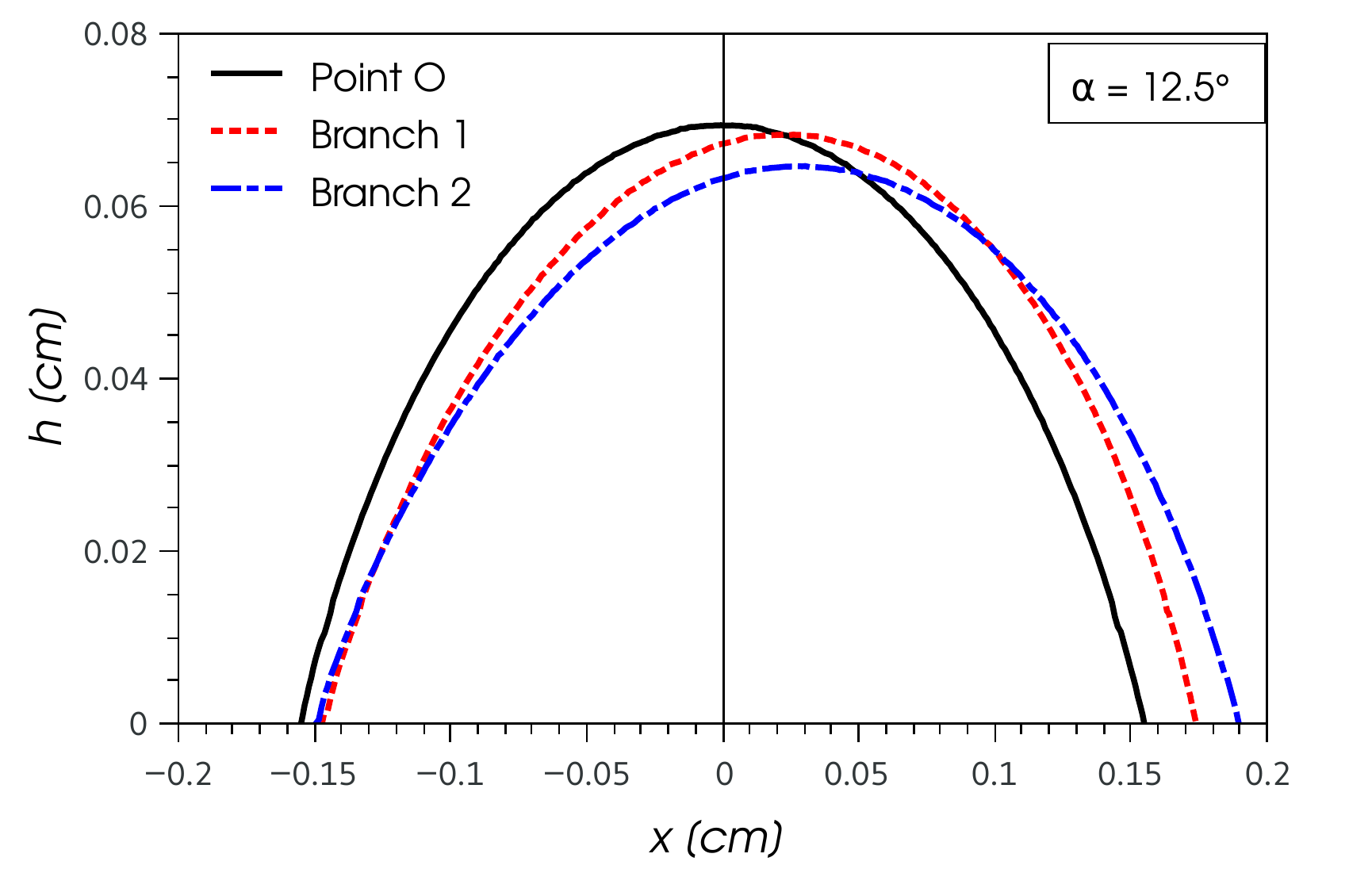}}
\subfigure[]{\includegraphics[width=0.42\textwidth]{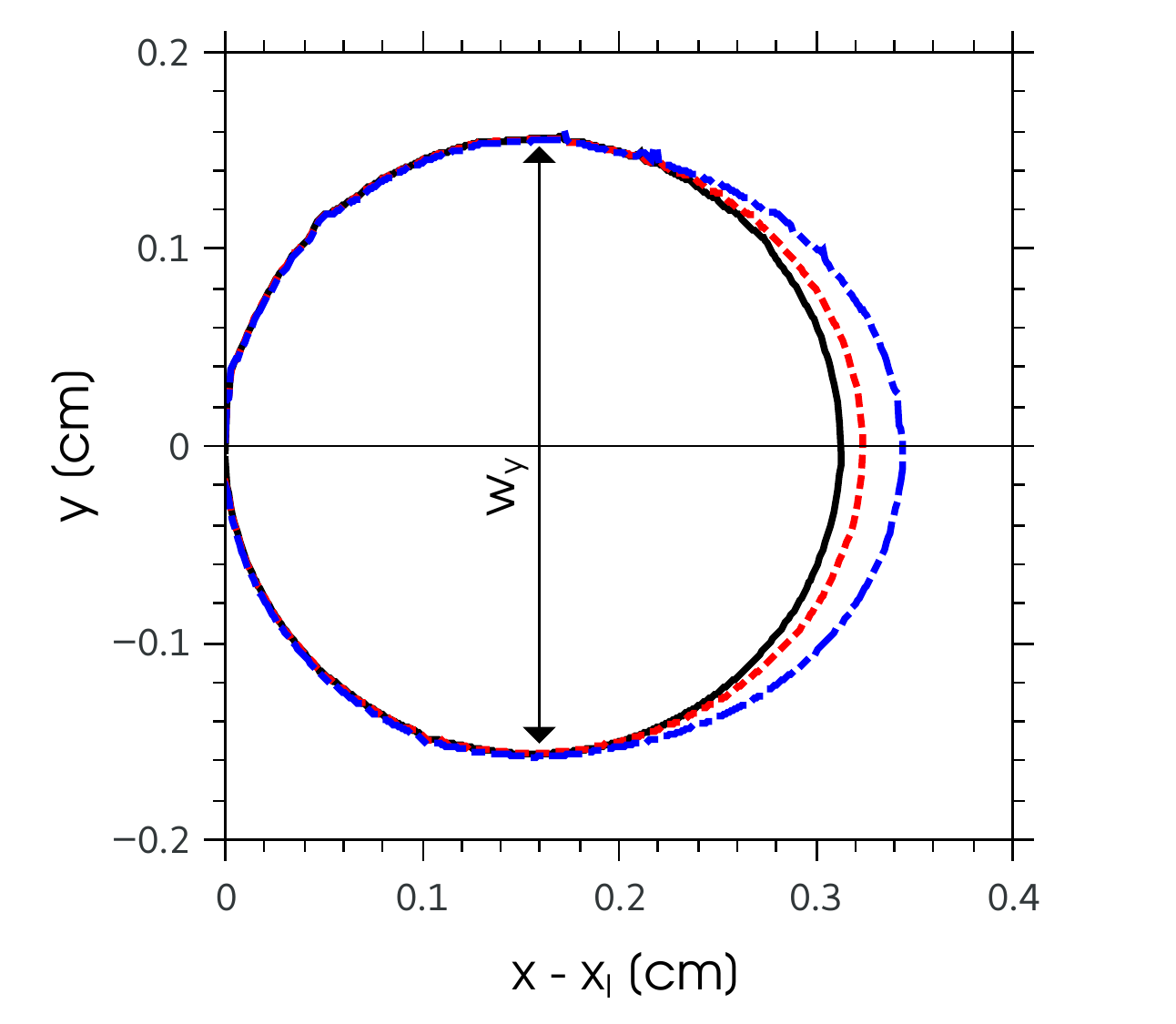}}
\caption{Thickness profiles for $\alpha=12.5^\circ$ at branch $1$ (dashed line)
and branch $2$ (dot--dashed lines). The solid line corresponds to point O at
$\alpha=0$ and it is shown for comparison. (b) Corresponding footprints shifted
so that the most left points are coincident.}
\label{fig:hx_alpha125_R12}
\end{figure}

As the plane is reverted to the horizontal position, the drop does not recover
the same shapes for a given $\alpha$ in different branches due to the hysteresis
of the contact angle. For instance, for $\alpha=12.5^\circ$ only the position
$x̣_l$ is almost coincident when comparing branches $1$ and $2$, but all other
parameters differ. The effects for the thickness profiles and footprints with
the same $\alpha$ are shown in Fig.~\ref{fig:hx_alpha125_R12} .

\begin{figure}[htb]
\centering
\subfigure[]{\includegraphics[width=0.45\textwidth]{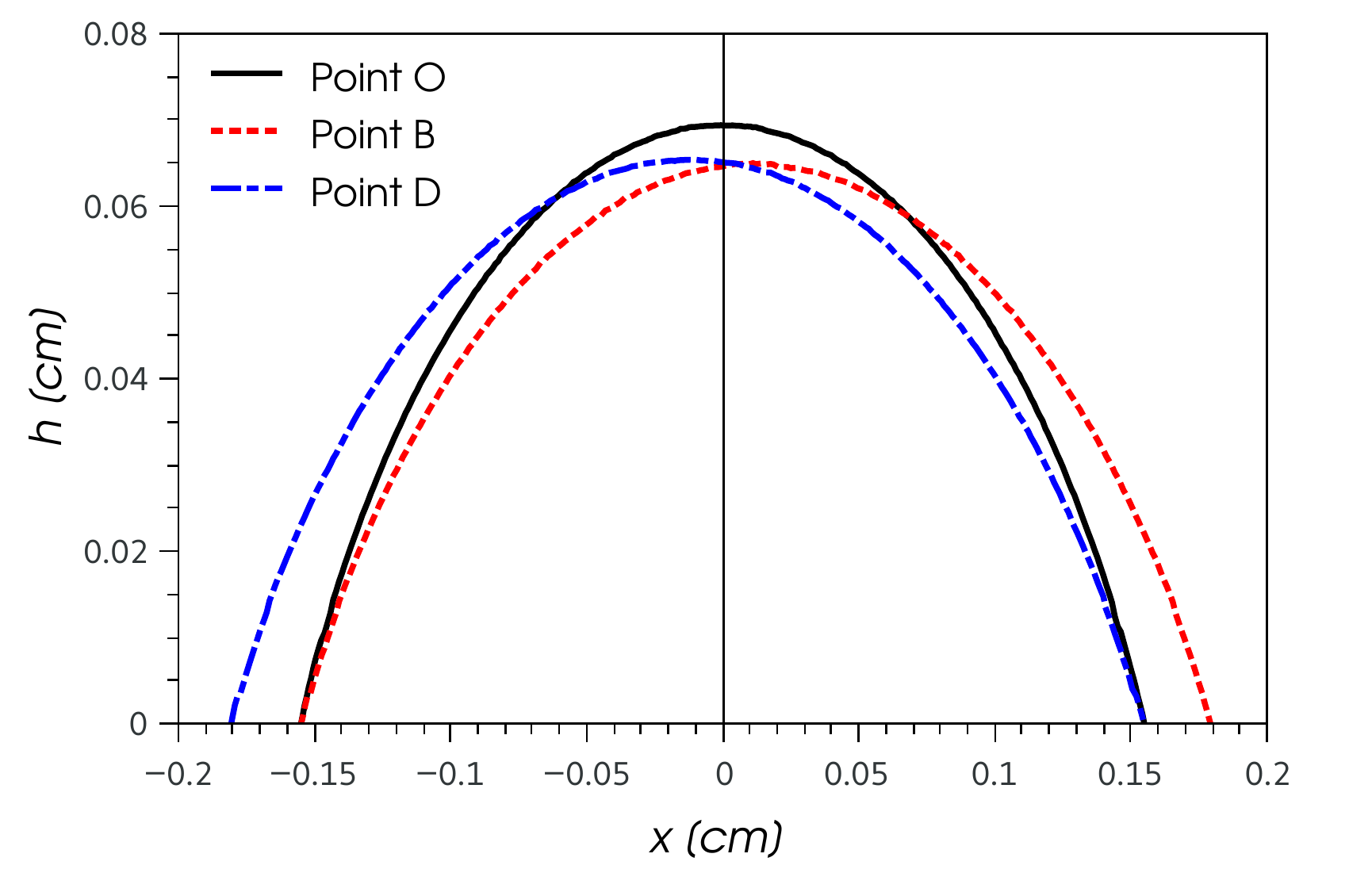}}
\subfigure[]{\includegraphics[width=0.45\textwidth]{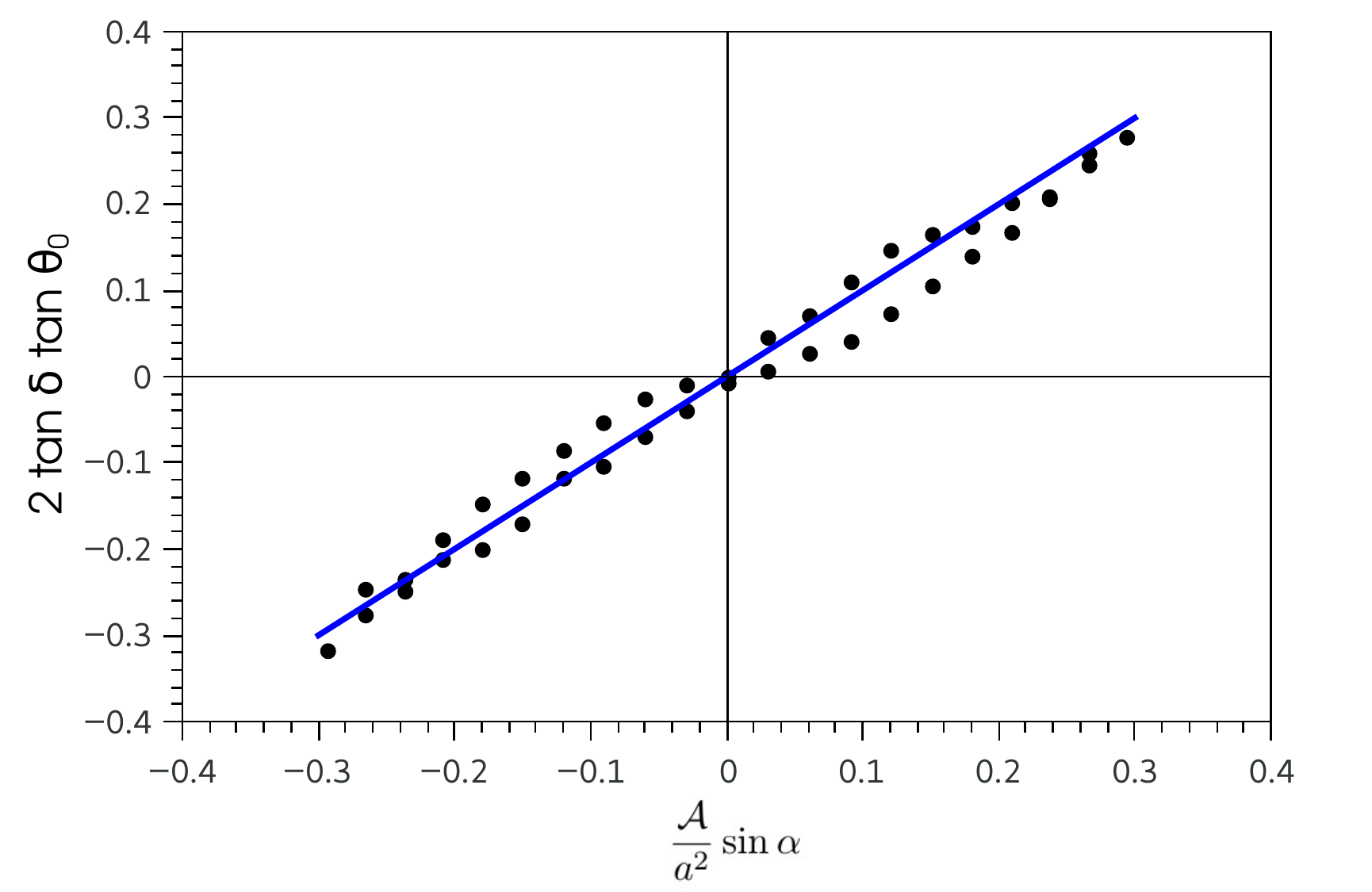}}
\caption{(a) Thickness profiles for $\alpha=0^\circ$ at different extreme points
of the cycle: Solid line at start (point O), dashed line at the end of branch
$2$ (point B), and dot--dashed line at the end of branch $4$ (point D).
(b) Experimental data plotted in terms of the left and right hand sides of
Eq.~(\ref{eq:tan_tan}) compared with the identity line.) 
Note that the profiles are practically coincident, if they are shifted towards
left and right, respectively, till both maximums are at $x=0$.}
\label{fig:hx_alpha0_A1d}
\end{figure}

We also verify that the profiles as well as the footprints depend not only on
$\alpha$ but also on the direction in which $\alpha$ is changing, i.e. whether
it is increasing or decreasing. In fact, we compare the profiles for $\alpha=12.5^\circ$ for branches 1 ($\alpha>0$) and 3 ($\alpha<0$) coincide when $|\alpha|$ is increasing.
The profiles for branches 2 ($\alpha>0$) and 4 ($\alpha<0$) are identical when
$|\alpha|$ is decreasing. This confirms the fact that (as expected) the sign of
$\alpha$ is irrelevant.

Fig~\ref{fig:hx_alpha0_A1d}a compares the profiles for
$\alpha=0$ at points O, B and D, where the drop has been affected by both
positive and negative inclination angles. Clearly, the profile at B and D do not
coincide with that at O, but those at B and D are coincident if they are shifted
to the left and right, respectively, till both maximums are at $x=0$. This fact
implies that the effect of changing the sign of $\alpha$ only has the effect of
exchanging the roles of the points $x_l$ and $x_r$, which is an expected result.
Therefore, this experimental verification reinforces the confidence in the
accuracy of our measurements.

When we consider the area of the thickness profile, ${\cal A}$, all along the cycle, we
find that it remains practically constant with a value ${\cal A}=(0.694 \pm 0.007)
a^2$. This allows us to compare this result with the predictions for a two
dimensional drop~\cite{krasovitski_lang05}, i.e. for the problem of an
infinitely long filament placed transversal to the incline. For instance,
in~\cite{dgk_pof12} we find the following relationship between ${\cal A}$ and the
contact angles,
\begin{equation}
 2 \tan \delta \tan \theta_0 = \frac{{\cal A}}{a^2} \sin \alpha
 \label{eq:tan_tan}
\end{equation}
where $\delta = (\theta_r-\theta_l)/2$ and $\theta_0=(\theta_r+\theta_l)/2$. In
order to compare this prediction with the present experiments, we plot the l.h.s
of Eq.~(\ref{eq:tan_tan}) as obtained from the measured contact angles versus
its r.h.s. considering the measured values of ${\cal A}$ (see symbols in
Fig.~\ref{fig:hx_alpha0_A1d}b). The good agreement with the line with the
identity line (representing Eq.~(\ref{eq:tan_tan})) shows that this formulation
is still valid for three dimensional drops as those studied in this work.

\section{Drop shape}
\label{sec:drop}

One main observation of the experiments reported in Section~\ref{sec:exp_data},
as well as in the literature (see
e.g.~\cite{extrand_jcis90,pierce_csA08,dussan_jfm85}), is that the shape of the
footprint does not remain circular for $\alpha>0$. However, since most of the
analytical approaches restrict to small values of $\alpha$, those studies have
considered a circular shape for the
footprint~\cite{elsherbini_jcis04,coninck_pre17}. Instead, we will extend now
our previous theory in~\cite{rava_pof16} for non--circular footprints on
horizontal planes to inclined ones. In particular, we will adapt that formalism
in polar coordinates to the present case.

The governing equation for the thickness profile of the static drop can be
obtained by considering the balance between the capillary pressure and both
components of the gravitational force. In dimensionless form, we have
\begin{equation}
 -\kappa + h \cos \alpha - x \sin \alpha = P=const.,
 \label{eq:full0}
\end{equation}
where the thickness $h(x,y)$ and the spatial coordinates $(x,y)$ are in units of
$a$, and $P$ is the drop pressure in units of $\gamma/a$. Here, $\kappa= \nabla
\cdot {\bf n}$ is the curvature of the drop free surface with normal vector
${\bf n} = \nabla F/|\nabla F |$, where $F = z - h(x,y) = 0$ defines it. Thus,
we have
\begin{equation}
 -\nabla \cdot \left( \frac{\nabla h}{\sqrt{1+ \epsilon |\nabla h|^2}} \right) + h \cos \alpha - x \sin \alpha = P=const.,
 \label{eq:full}
\end{equation}
For $\epsilon=1$, the first term stands for the full surface curvature, while for $\epsilon=0$ it yields the curvature only valid for small free surface slopes, i.e. $|\nabla h|^2\ll 1$, in the context of the long--wave theory (lubrication approximation). The solution domain is the drop footprint, whose
 border line is denoted by a closed curve $\Gamma_{\alpha}(x,y)=0$ where $h=0$ for a given $\alpha$. The determination of the constant $P$ and the footprint shape, $\Gamma_{\alpha}(x,y)=0$, depends on the approach used to solve Eq.~(\ref{eq:full}) (see Sections~\ref{sec:num} and \ref{sec:lubt}).

A first approach to simplify this equation consists on assuming the validity of
the lubrication approximation, even if the drops considered here do not have
small contact angles. Thus, Eq.~(\ref{eq:full}) with $\epsilon=0$ reads as (see
e.g.~\cite{dgk_pof12})
\begin{equation}
 -\nabla^2 h + h \cos \alpha - x \sin \alpha =P,
 \label{eq:lub}
\end{equation}
which is a linear equation, similar to that one studied in~\cite{rava_pof16}. In
fact, its solution can be written in the form  
\begin{equation}
 h= h_1 + \frac{P}{\cos \alpha} + x \tan \alpha
\end{equation}
where $h_1$ satisfies the homogeneous equation
\begin{equation}
 \nabla^2 h_1 - h_1 \cos \alpha=0.
 \label{eq:homog}
\end{equation}
In order to obtain $h_1$, it is convenient to define the polar coordinates
\begin{equation}
 r=\sqrt{{x}^2 + y^2}, \qquad \varphi = \arctan \frac{y}{x},
 \label{eq:rphi}
\end{equation}
where $x$--origin is defined at the $x$--coordinate of the point on
$\Gamma_{\alpha}(x,y)$ with maximum $y$. Thus, this point corresponds to $(0,w_y/2)$,
where $w_y$ is the drop width in the transverse direction. Within this reference
frame, we assume a factorized solution of Eq.~(\ref{eq:homog}) as $h_1=R(r) \Phi(\varphi)$, and find
\begin{equation}
h(r, \varphi) = \frac{P}{cos \, \alpha} + r \, cos\, \varphi \, tan \, \alpha  + \sum_{m=0}^{\infty}(A_m \, cos\, m \varphi + B_m\, sin\, m \varphi) I_m(r), 
\label{eq:hseries}
\end{equation}
where $I_m(r)$ is the modified Bessel function of the first kind. In order to
have a symmetric solution with respect to the $x$--axis, we must have $B_m = 0$
for all $m$, and then the problem reduces to finding the remaining constants
$A_m$. In principle, this could be accomplished by setting $h=0$ at an infinite
number of points along the footprint border, $\Gamma_{\alpha}(x,y)=0$. The evaluation
of this series should lead to the same free surface shape as the numerical
solution of Eq.~(\ref{eq:lub}). In Section~\ref{sec:lubt}, we will develop an
approximation of this series to make this solution of practical use.

\section{Numerical approach and comparison with experiments}
\label{sec:num}

In this section, we numerically solve Eqs.~(\ref{eq:full}) and (\ref{eq:lub}).
The integration domain $\Gamma_{\alpha}(x,y)$ is obtained by means of an
interpolation curve which fits the footprint data from the experiments. The
constant $P$ is adjusted to a given drop volume, $V$, by means of an iterative
procedure.

\begin{figure}[htb]
\centering
\subfigure[]{\includegraphics[width=0.329\textwidth]{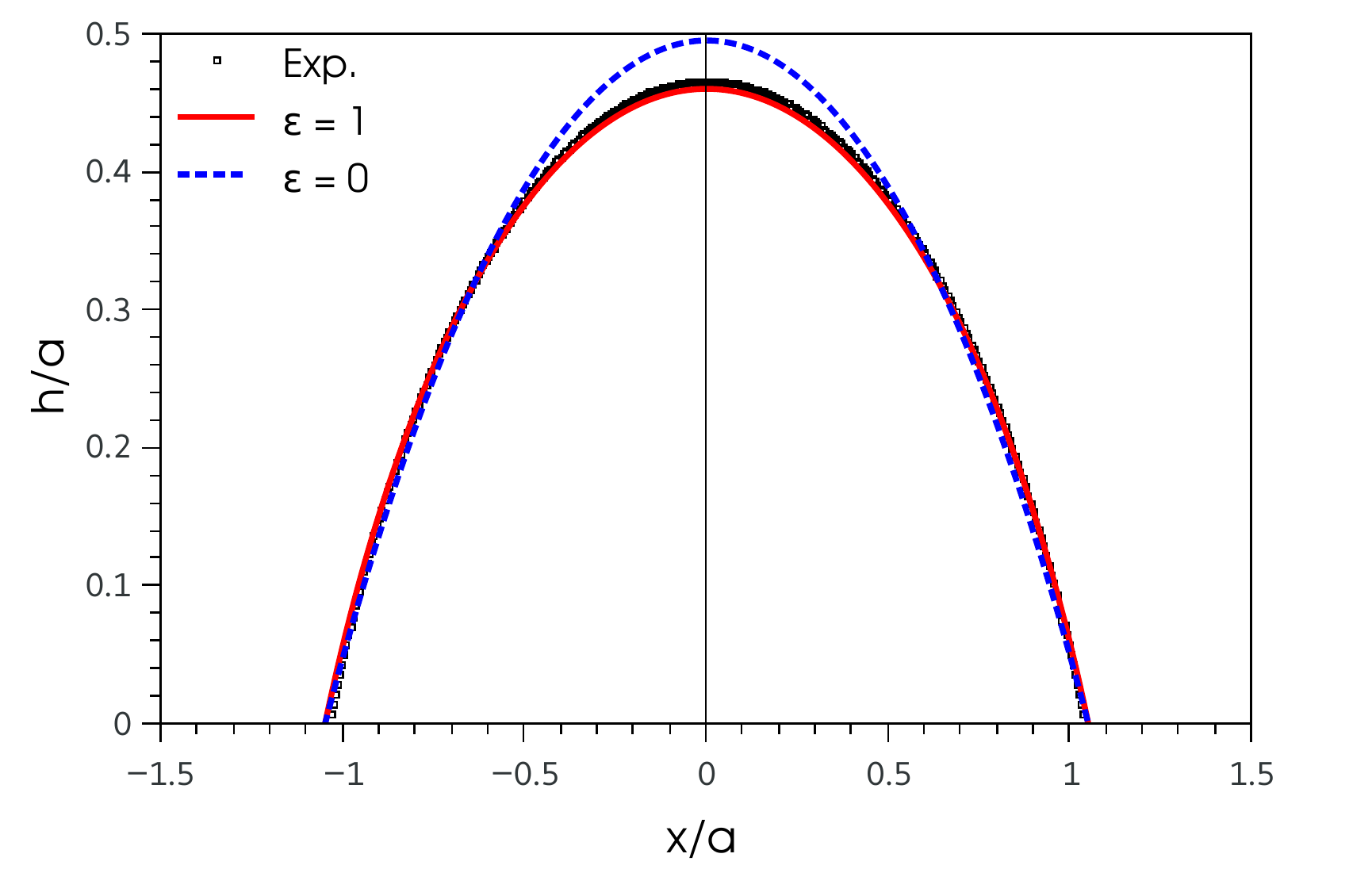}}
\subfigure[]{\includegraphics[width=0.329\textwidth]{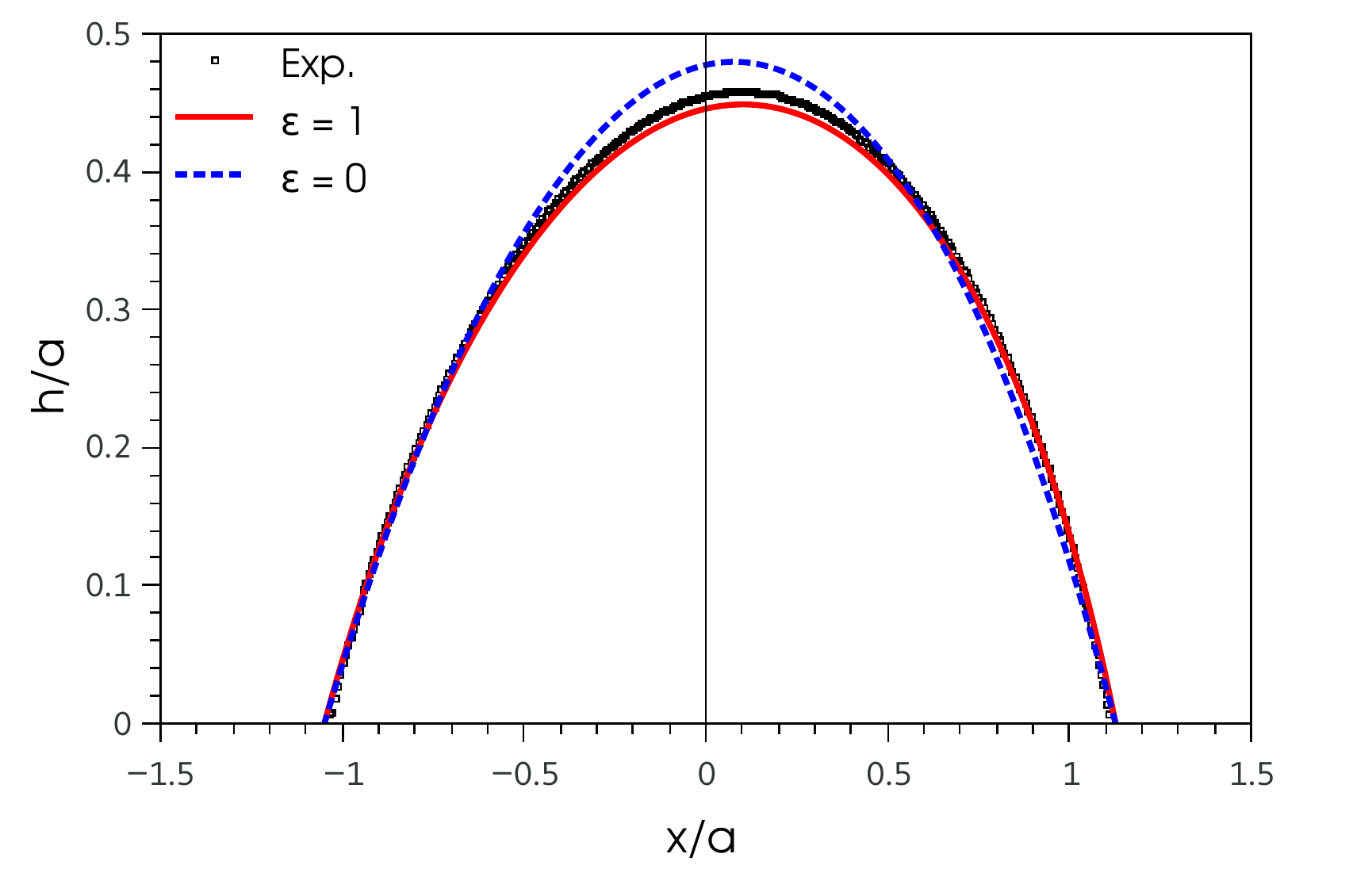}}
\subfigure[]{\includegraphics[width=0.329\textwidth]{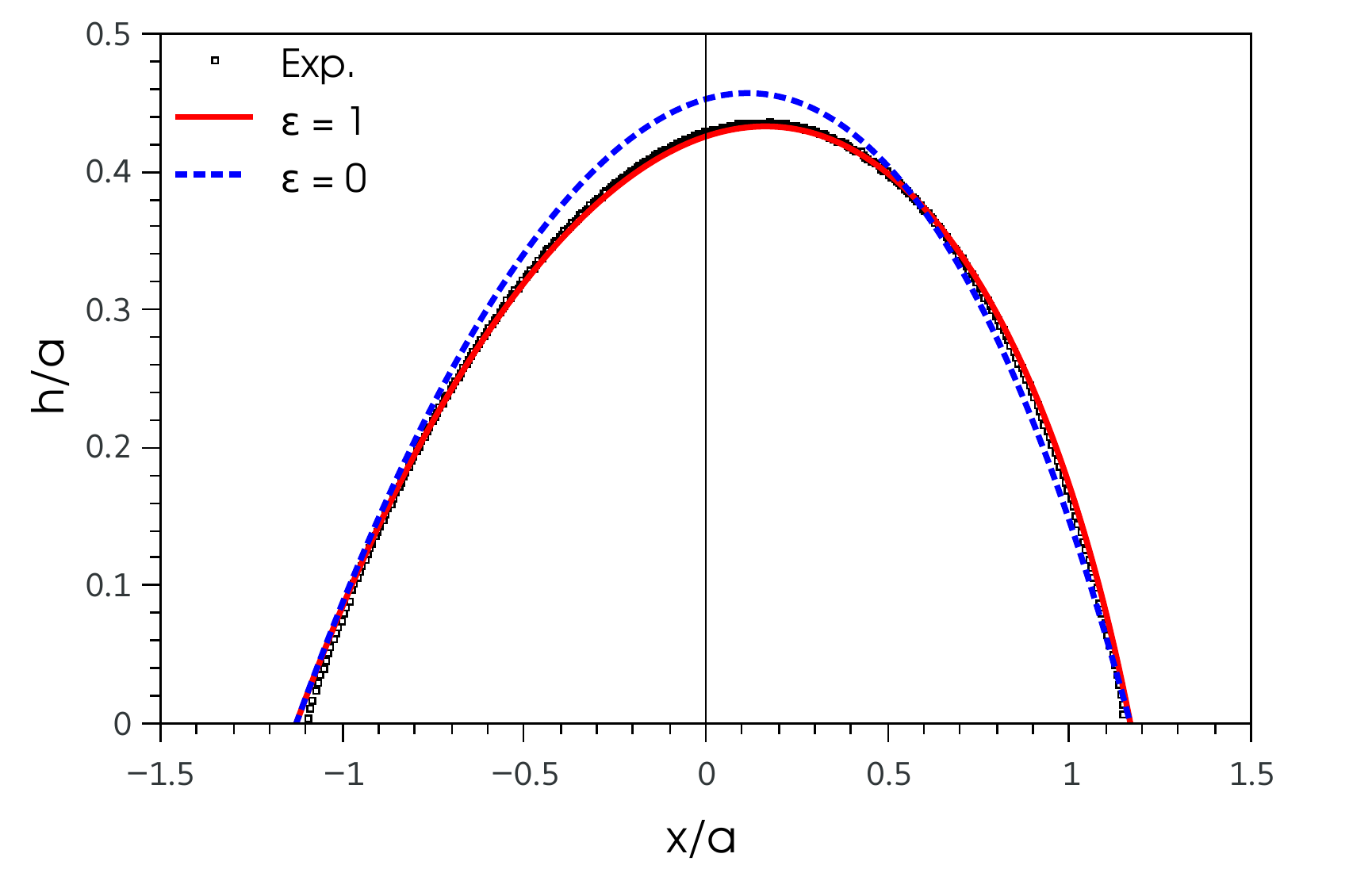}}
\caption{Comparison between the experimental thickness profile (symbols) and the
full ($\epsilon=1$) and lubrication approximation ($\epsilon=0$) solutions. (a) $\alpha=0^\circ$, (b) $\alpha=12.5^\circ$, and (c) $\alpha=25^\circ$}
\label{fig:hx_exp_num}
\end{figure}

The value of $V$ is that one measured for $\alpha=0$ from the thickness profile
assuming axial symmetry, i.e. a circular footprint. Figure~\ref{fig:hx_exp_num}
shows a comparison of the longitudinal numerical thickness profiles, $h(x,y=0)$,
with the experimental ones (symbols) for $\alpha=0^\circ$, $12.5^\circ$ and
$25^\circ$. Clearly, the full equation for $\epsilon=1$ fits better these
profiles than the long--wave approximation for $\epsilon=0$. Since the three
dimensional solutions have the same volume as the experimental drop,
compensations of thickness can be seen along other longitudinal or cross
sections. For brevity, we do not show here these differences.

\begin{figure}[htb]
\centering
\subfigure[]{\includegraphics[width=0.329\textwidth]{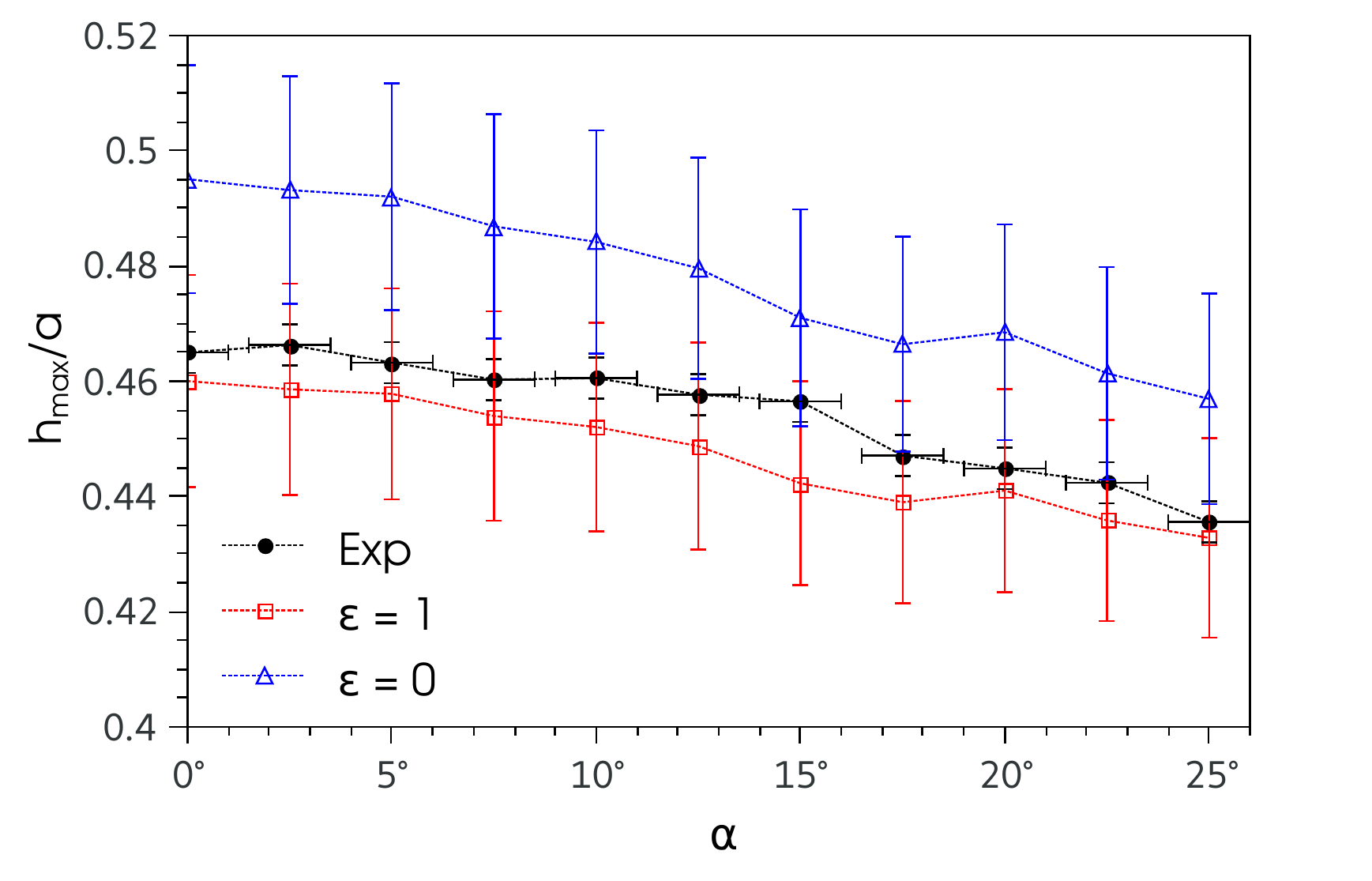}}
\subfigure[]{\includegraphics[width=0.329\textwidth]{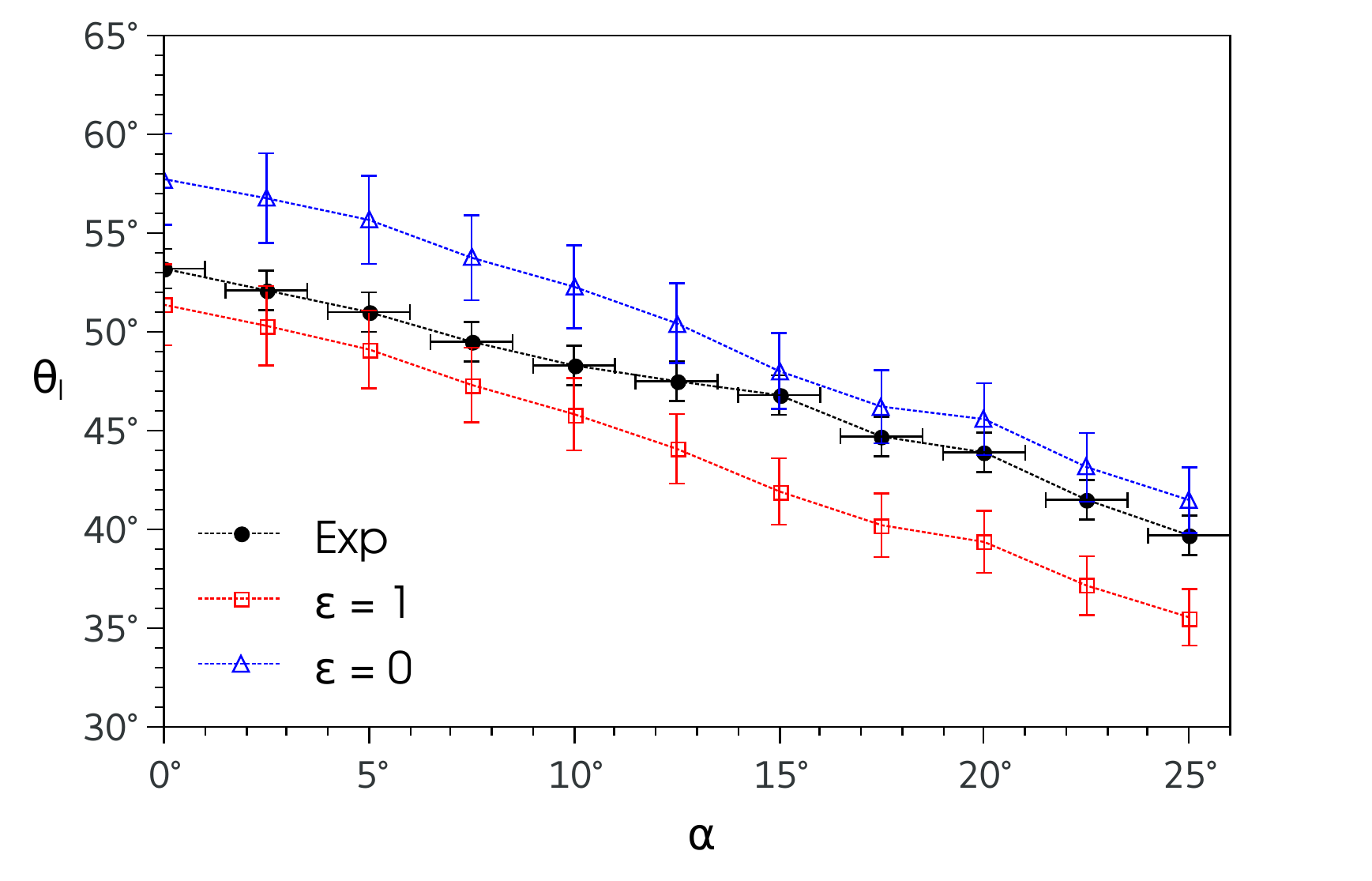}}
\subfigure[]{\includegraphics[width=0.329\textwidth]{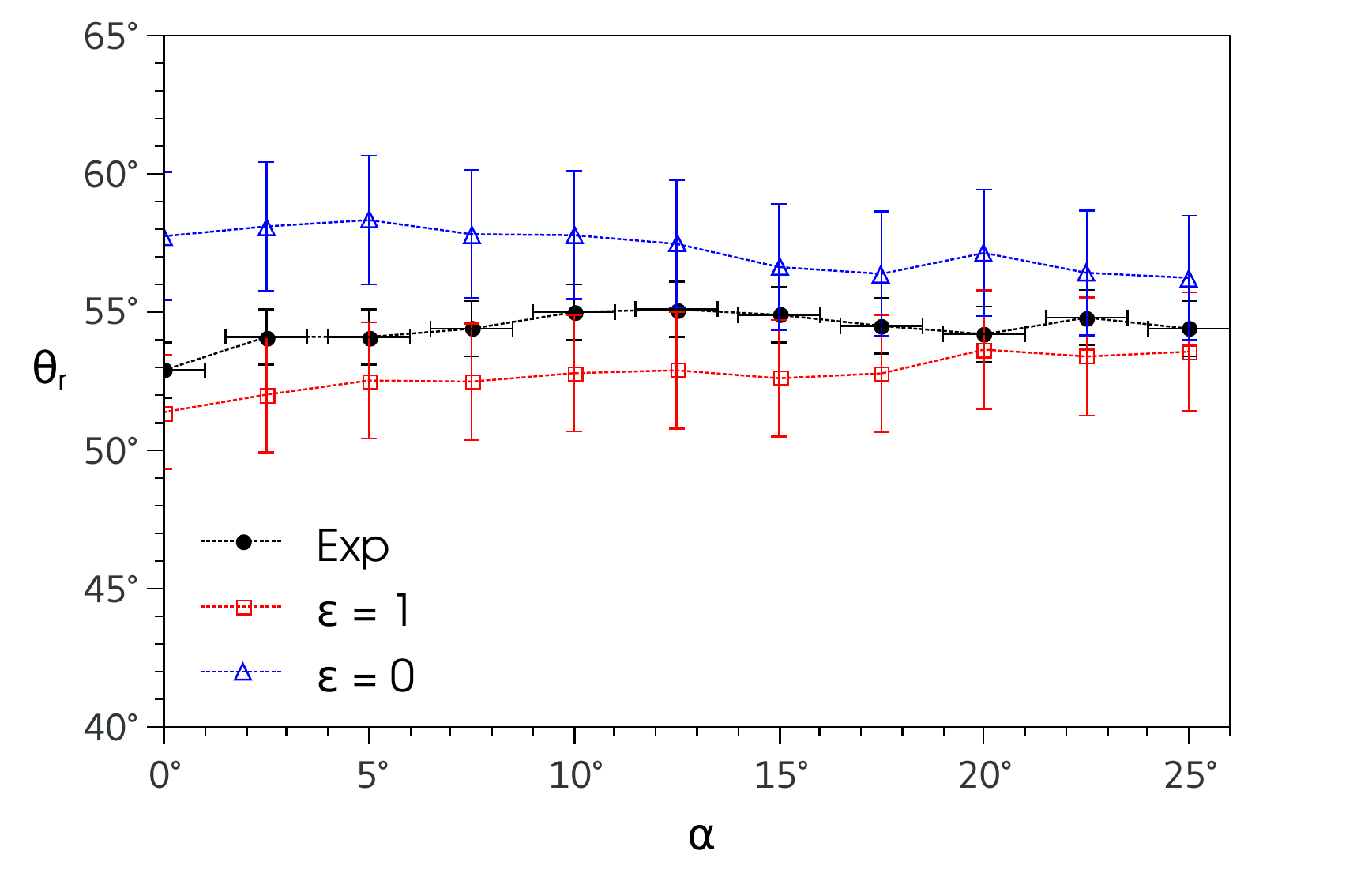}}
\caption{Comparison between the experimental data and the numerical solutions
for $\epsilon=1$ and $\epsilon=0$ (Eqs.~(\ref{eq:full}) and (\ref{eq:lub})) at
different inclination angles, $\alpha$. Both the footprint contour and the drop
volume, $V=(0.876 \pm 0.044)a^3=(2.9 \pm 0.14) mm^3$, are given for each calculation. (a) Maximum
thickness of the drop profile, $h_{max}$. (b) Contact angle at the left side,
$\theta_l$. (c) Contact angle at the right side, $\theta_r$. }
\label{fig:Exp_Num}
\end{figure}

The relative error $e_V=\Delta V / V$ can be estimated from the corresponding
errors in the measurement of $h_{max}$ and $x_r$ (or $x_l$) which are $3\%$ and
$1\%$, respectively, so that $e_V=5\%$. Consequently, the numerical solutions
are also affected by $e_V$, since their predictions (such as thickness, contact
angles, etc.) depend on $V$ through the iterative determination of the constant
$P$ in Eqs.~(\ref{eq:full}) and (\ref{eq:lub}). We have estimated this
propagation in about $3\%$. Besides, we have also considered the error of $\pm
1^\circ$ in the measurement of $\alpha$. In Fig.~\ref{fig:Exp_Num}a we show the
comparison of $h_{max}$ with the experimental data for the whole range of
$\alpha$, along with the corresponding error bars.

Regarding the calculation of the contact angles for the thickness profile
$h(x,y=0)$ at $x_l$ and $x_r$, we must be careful to use the correct order of
approximation for both $\epsilon=0,1$. This is because the approximation $\tan
\theta \approx \theta$ is used for $\epsilon=0$, and in general, $\theta$ is not
so small as to fully satisfy this simplification. Therefore, we write the
general formulae,
\begin{eqnarray}
 \theta (x_r,y=0)&=&  -\epsilon \arctan \left( \frac{\partial h}{\partial x} \right) - 
              (1 -\epsilon) \left( \frac{\partial h}{\partial x} \right),\nonumber \\ 
 \theta (x_l,y=0)&=&  \epsilon \arctan \left( \frac{\partial h}{\partial x} \right) +
              (1 -\epsilon) \left( \frac{\partial h}{\partial x} \right).
\end{eqnarray}
Note that $\arctan (\partial h /\partial x) > (\partial h /\partial x)$ for
$(\partial h /\partial x) \gtrsim 1$. The comparison with the experimental data
is shown in Fig.~\ref{fig:Exp_Num}b and c.

In summary, these results allow us to state that the agreement of the numerical
solution for $\epsilon=0$ with the experimental data is marginally within the
errors, and that this approximation always yields overestimated values.

\subsection{Contact angle distribution along the drop periphery}
\label{sec:cangle}

An important result of the numerical approaches is the contact angle
distribution around the drop periphery, i.e the function $\theta(\varphi)$. 
Except for the work in~\cite{elsherbini_jcis04}, where $\theta$ was measured at eight 
different azimuthal angles, only the contact angles at the points $(x_l,0)$ 
and $(x_r,0)$ of the $h(x,0)$ profile are usually reported in the experiments. 
The knowledge of this function has been the object of several previous works~\cite{elsherbini_jcis04,coninck_pre17}. Other authors have also performed numerical simulations to obtain $\theta(\varphi)$. Some have assumed a circular shape
for the footprint~\cite{brown_jcis80,milinazzo_jcis88} and others have used
non--circular shapes obtained by minimization of the required hysteresis
range~\cite{dimitrakopoulos_jfm99}. In general, all the functions
$\theta(\varphi)$, except that in~\cite{dimitrakopoulos_jfm99} which resembles a
step function with a linear transition region, show a smooth variation between $\theta_r$ and $\theta_l$~\cite{anapragada_ijhmt11}. 

The theoretical studies are limited to small Bond numbers, i.e. either small
$\alpha$ or small drop diameter, or both (see Eq.~(\ref{eq:Bond})). Instead, the
two numerical approaches presented here allow to obtain $\theta(\varphi)$
without any restriction on the value of the Bond number. 

To do so, we consider the $z$--component of the normal versor to the free
surface, and obtain
\begin{equation}
 n_z=\cos \theta(r(\varphi),\varphi) = \left[ 1 +\left( \frac{\partial h}{\partial r} \right)^2 + \left( \frac{1}{r}  \frac{\partial h}{\partial \varphi} \right)^2 \right]^{-1/2},\qquad \epsilon=1,
 \label{eq:thec_e1}
\end{equation}
where $r(\varphi)$ is the footprint boundary given by $\Gamma_{\alpha}(r,\varphi)=0$.
Within the long-wave approximation, we consider this equation for $\cos \theta
\approx 1 - \theta^2 / 2 $, and thus we find
\begin{equation}
\theta(r(\varphi),\varphi)= \pm \sqrt{ \left( \frac{\partial h}{\partial r} \right)^2 + \left( \frac{1}{r}  \frac{\partial h}{\partial \varphi} \right)^2 },\qquad \epsilon=0.
\label{eq:thec_e0}
\end{equation}

\begin{figure}[htb]
\centering
\subfigure[]{\includegraphics[width=0.45\textwidth]{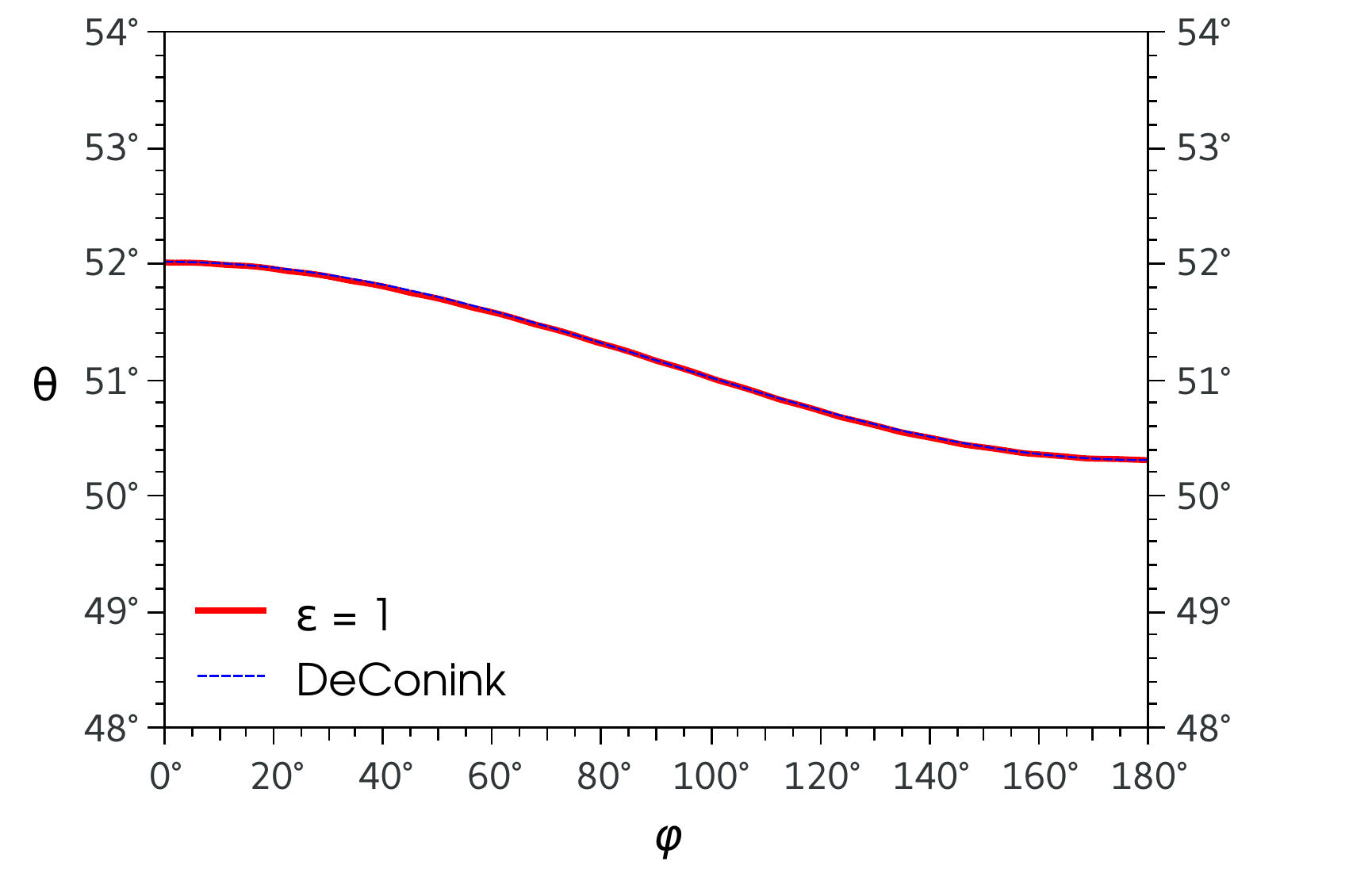}}
\subfigure[]{\includegraphics[width=0.45\textwidth]{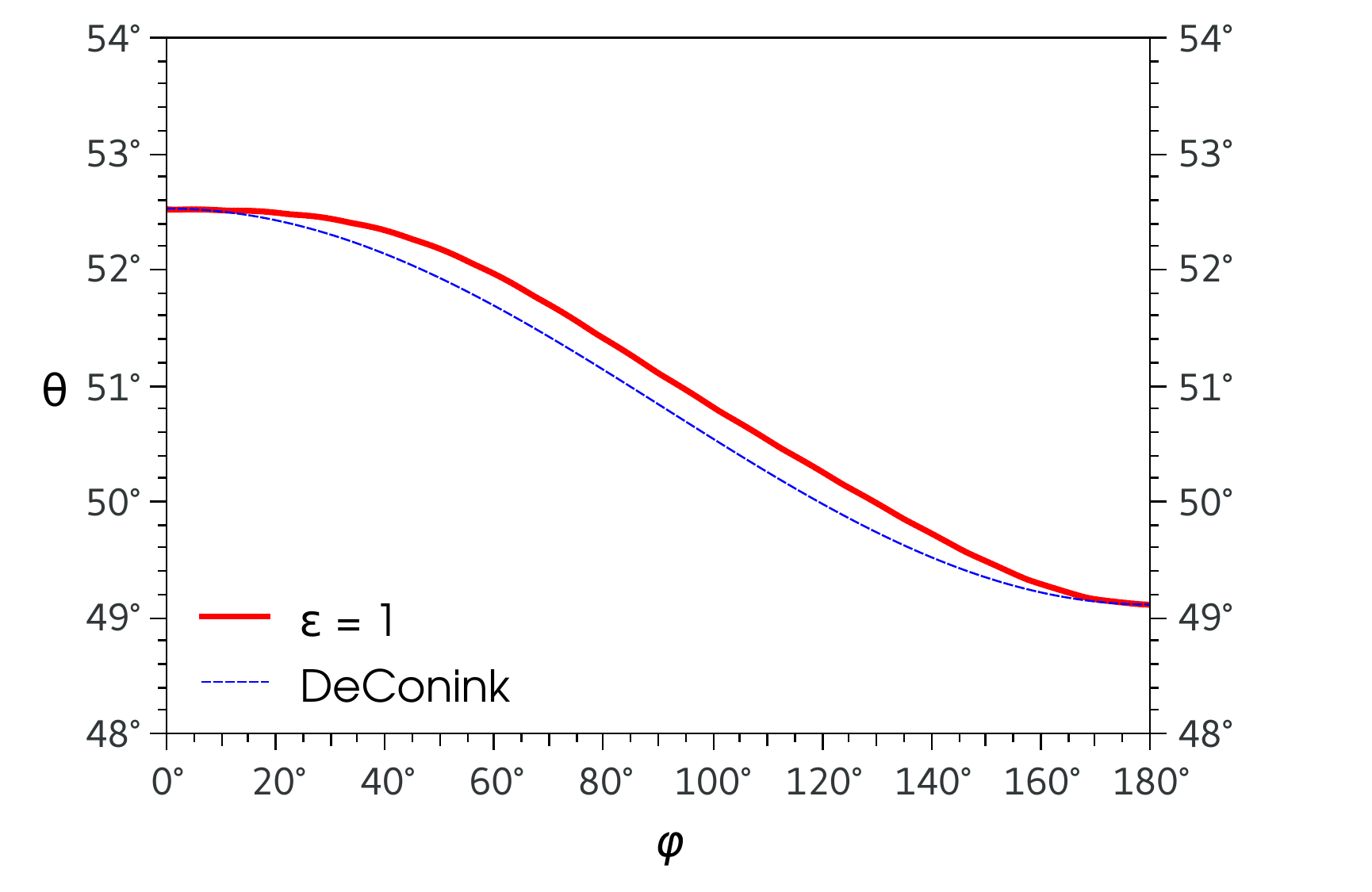}}
\subfigure[]{\includegraphics[width=0.45\textwidth]{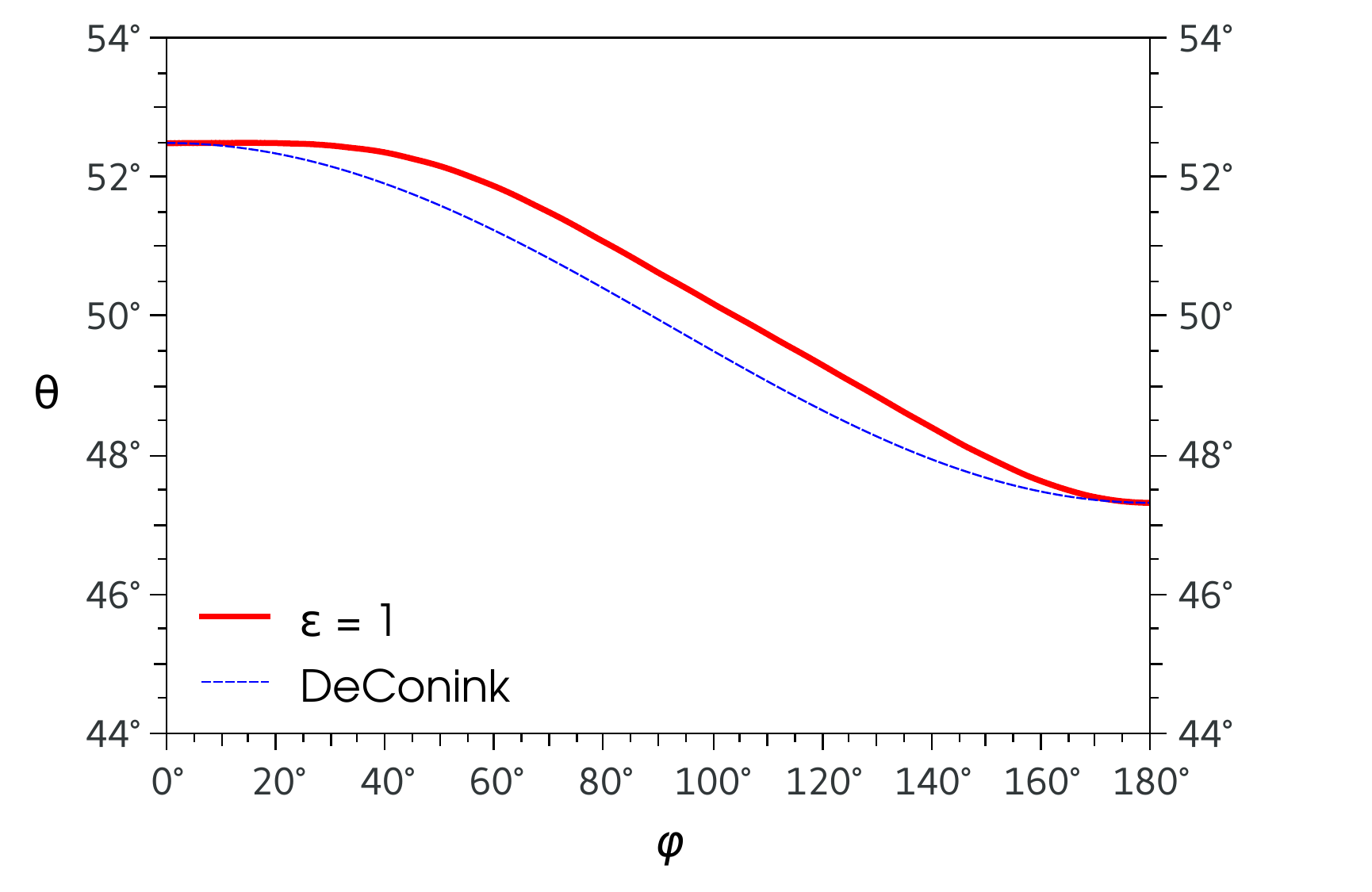}}
\subfigure[]{\includegraphics[width=0.45\textwidth]{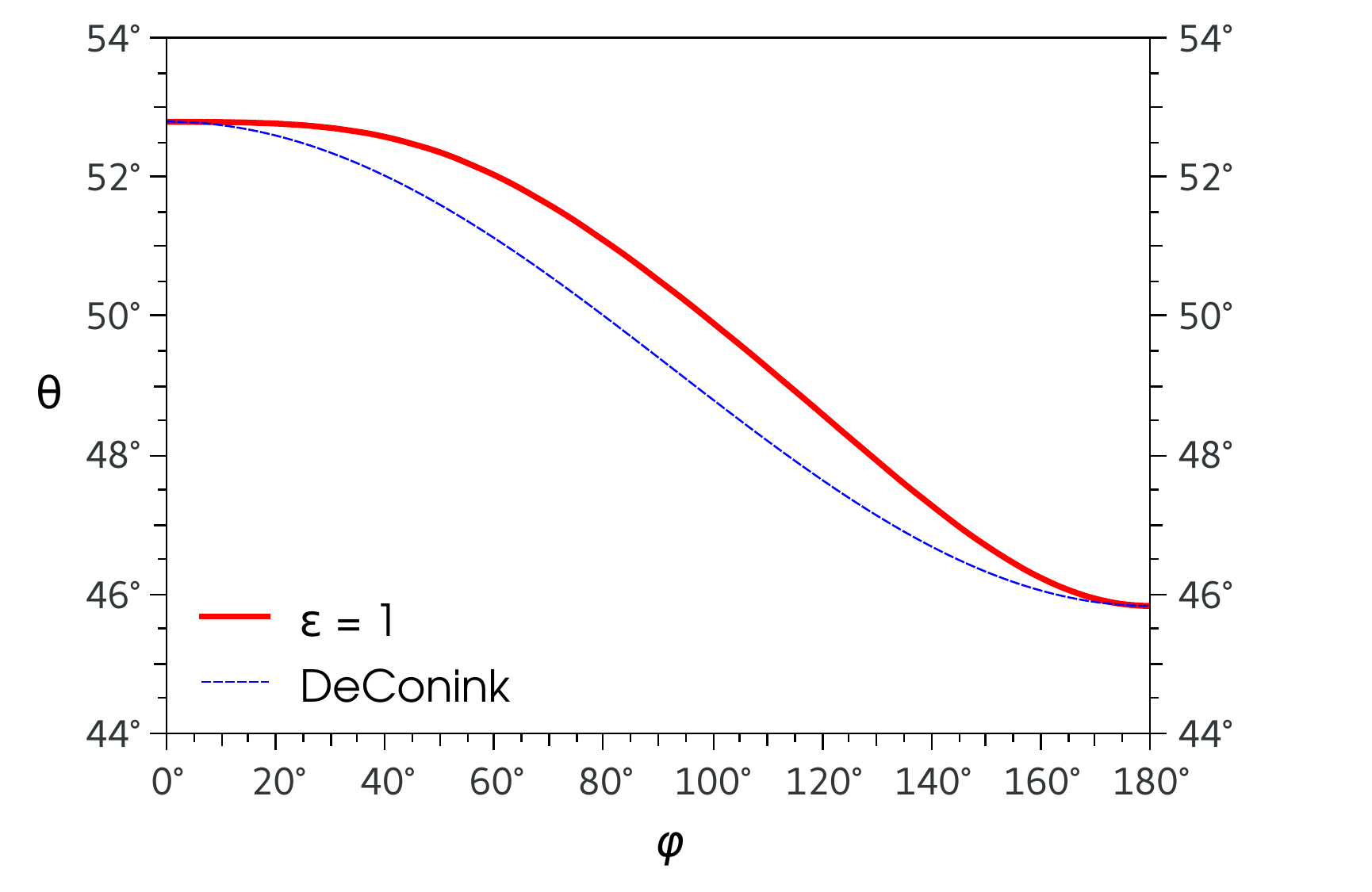}}
\caption{Azimuthal distribution of contact angle, $\theta$, for various inclination angles, $\alpha$: (a) $\alpha=2.5^\circ$ ($Bo=0.19$, $\theta_r=52.0^\circ$, $\theta_l=50.3^\circ$), (b) $\alpha=5^\circ$ ($Bo=0.38$, $\theta_r=52.5^\circ$, $\theta_l=49.1^\circ$), (c) $\alpha=7.5^\circ$ ($Bo=0.58$, $\theta_r=52.0^\circ$, $\theta_l=47.3^\circ$), and (d) $\alpha=10^\circ$ ($Bo=0.79$, $\theta_r=52.8^\circ$, $\theta_l=45.8^\circ$). The values at $\varphi=0, \pi$ stand for the extreme footprint points facing downslope and upslope, respectively. The thick solid red line corresponds to the solution of Eq.~(\ref{eq:full}). The dashed lines correspond to Eq.~(\ref{eq:conick}).}
\label{fig:the_full_ref}
\end{figure}

The prediction for $\epsilon=1$ is compared with the small $Bo$ theory developed by De
Coninck et al.~\cite{coninck_pre17}, which is developed without the lubrication approximation. This solution for $\theta(\varphi)$ is written in the form
\begin{equation}
 \cos \theta = \left( \cos \theta_l - \cos \theta_r \right) \frac{1-\cos \varphi}{2} + \cos \theta_r,
 \label{eq:conick}
\end{equation}
where $\theta_r$ and $\theta_l$ are given values.
The comparison is done with the values $\theta_r$ and $\theta_l$ as obtained from the numerical solution for $\epsilon=1$ at different $\alpha$'s by using the experimental footprints (see Fig.~\ref{fig:the_full_ref} for $\alpha=2.5^\circ$, $5^\circ$,
$7.5^\circ$ and $10^\circ$). For $\alpha$ as small as $2.5^\circ$, the results are in very good agreement with the theory, thus probing that our numerical solution is correct. For larger $\alpha$'s, we obtain larger values of $\theta$ than predicted by Eq.~(\ref{eq:conick}), which is only valid for small $\alpha$'s.

\begin{figure}[htb]
\centering
\subfigure[]{\includegraphics[width=0.45\textwidth]{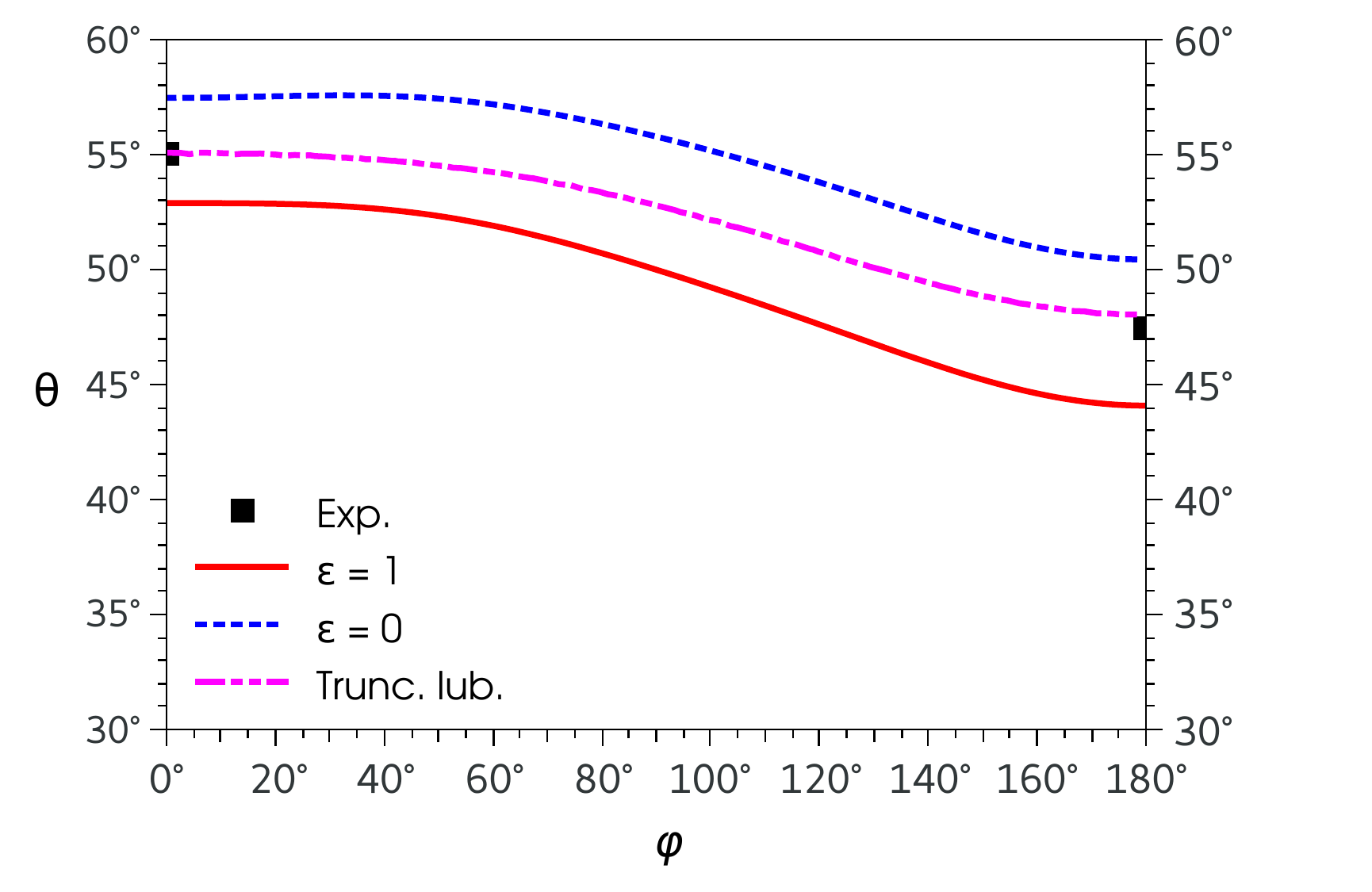}}
\subfigure[]{\includegraphics[width=0.45\textwidth]{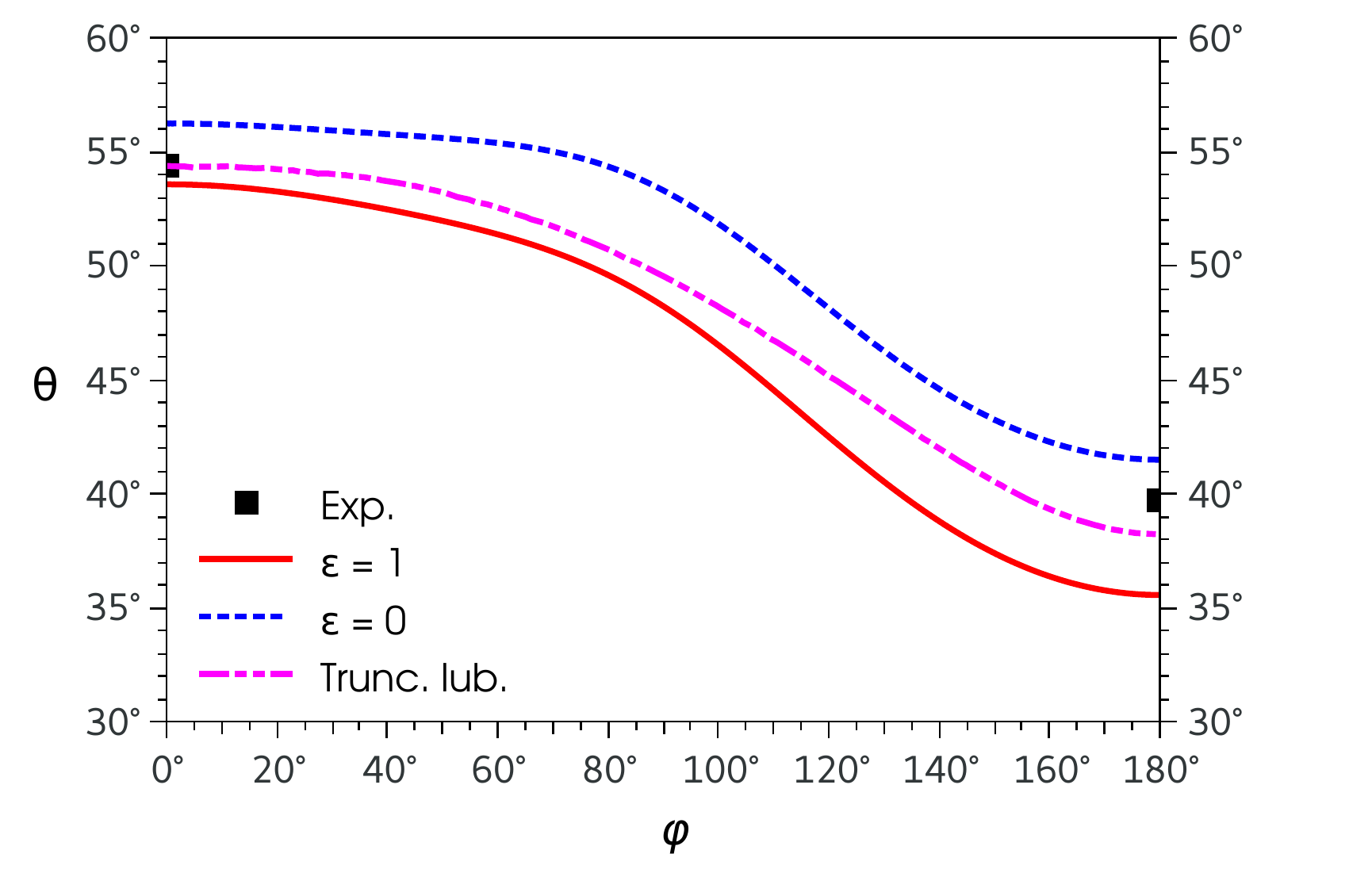}}
\caption{Comparison between the numerical solutions $\epsilon=1$ (red solid lines) and $\epsilon=0$ (blue dashed lines) for the azimuthal distribution of contact angle, $\theta$, for two large values of $\alpha$. (a) $\alpha=12.5^\circ$ ($Bo=1.01$), and (b) $\alpha=25^\circ$ ($Bo=2.15$).  The (magenta) dot--dashed lines correspond to the truncated analytical solution of the lubrication approximation. The two symbols at $\varphi=0^\circ$ and $180^\circ$ stand for the experimental data, namely, $\theta_l$ and $\theta_r$.}
\label{fig:the_num_exp}
\end{figure}

For even larger values of $\alpha$, we can compare the predictions with
$\epsilon=0$ and $1$ between them, as well as with the measured values of
$\theta$ at the most extreme points, i.e. $\theta_l$ and $\theta_r$ (see the
symbols in Fig.~\ref{fig:the_num_exp}a and b for $\alpha=12.5^\circ$ and
$25^\circ$). At $\varphi=0,\pi$, these experimental values are in between the
predictions for $\epsilon=0$ and $1$. For $0<\varphi<\pi$, we observe that for
larger $\alpha$ the size of the footprint region facing downhill ($ |\varphi |
\approx 0$, in which $\theta \approx \theta_r$) increases, and that the
transition from $\theta_l$ to $\theta_r$ becomes more and more abrupt.

\section{Analytical approach and comparison with experiments}
\label{sec:lubt}

In order to look for a useful analytical solution of Eq.~(\ref{eq:lub}), we
further assume that the shape of the drop can be reasonably estimated by the
first three terms of the series in Eq.~(\ref{eq:hseries}), so that
\begin{equation}
h(r, \varphi) \approx h_{trunc}(r, \varphi) =
\frac{P}{\cos \alpha} + r \, \tan \alpha \, \cos \varphi  +  A_0 \, I_0(r) + 
A_1 \, I_1(r) \, \cos \varphi + A_2 \, I_2(r) \, \cos 2 \varphi .
\label{eq:htrunc}
\end{equation}
The advantage of this approach is that the determination of the three constants
$A_i$ ($i=0,1,2$) requires to consider only three points at the footprint
border, and not the whole footprint as required in the numerical approach. Only
an infinite number of points in the footprint could determine all the series
coefficients. 

Here, we choose the points $(x=x_l,y=0)$, $(x=x_r,y=0)$, and $(x=0,y=w_y/2)$,
whose values are given as experimental data. Thus, we write the three conditions
in polar coordinates as
\begin{equation}
h_{trunc}(r=r_l,\varphi=\pi)= h_{trunc}(r=r_r,\varphi=0)=h_{trunc}(r=w_y/2,\varphi=-\pi/2) = 0.
\label{eq:cond_h}
\end{equation}
where $r_l=-x_l$ and $r_r=x_r$. 

\begin{figure}[htb]
\centering
\subfigure[]{\includegraphics[width=0.329\textwidth]{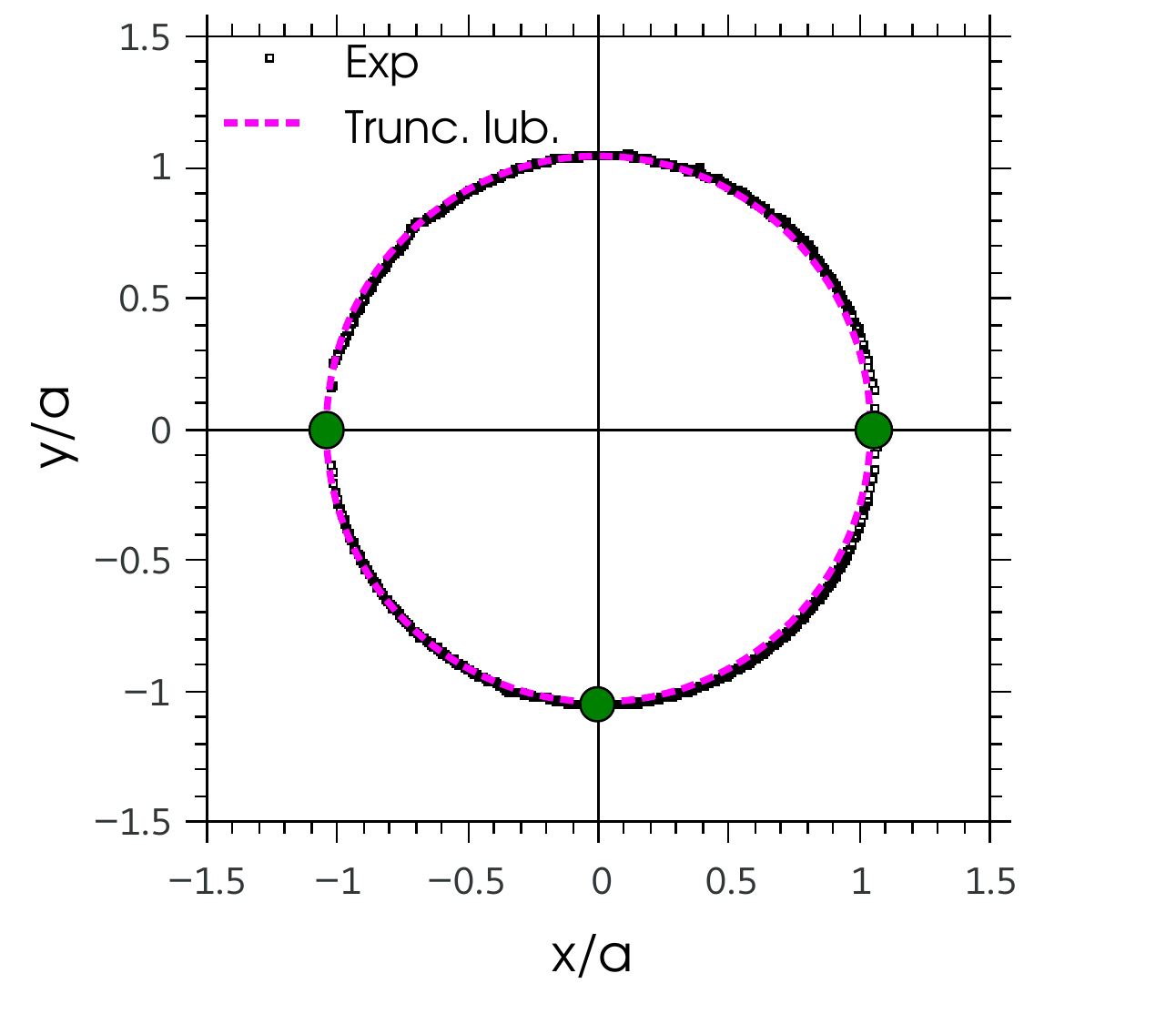}}
\subfigure[]{\includegraphics[width=0.329\textwidth]{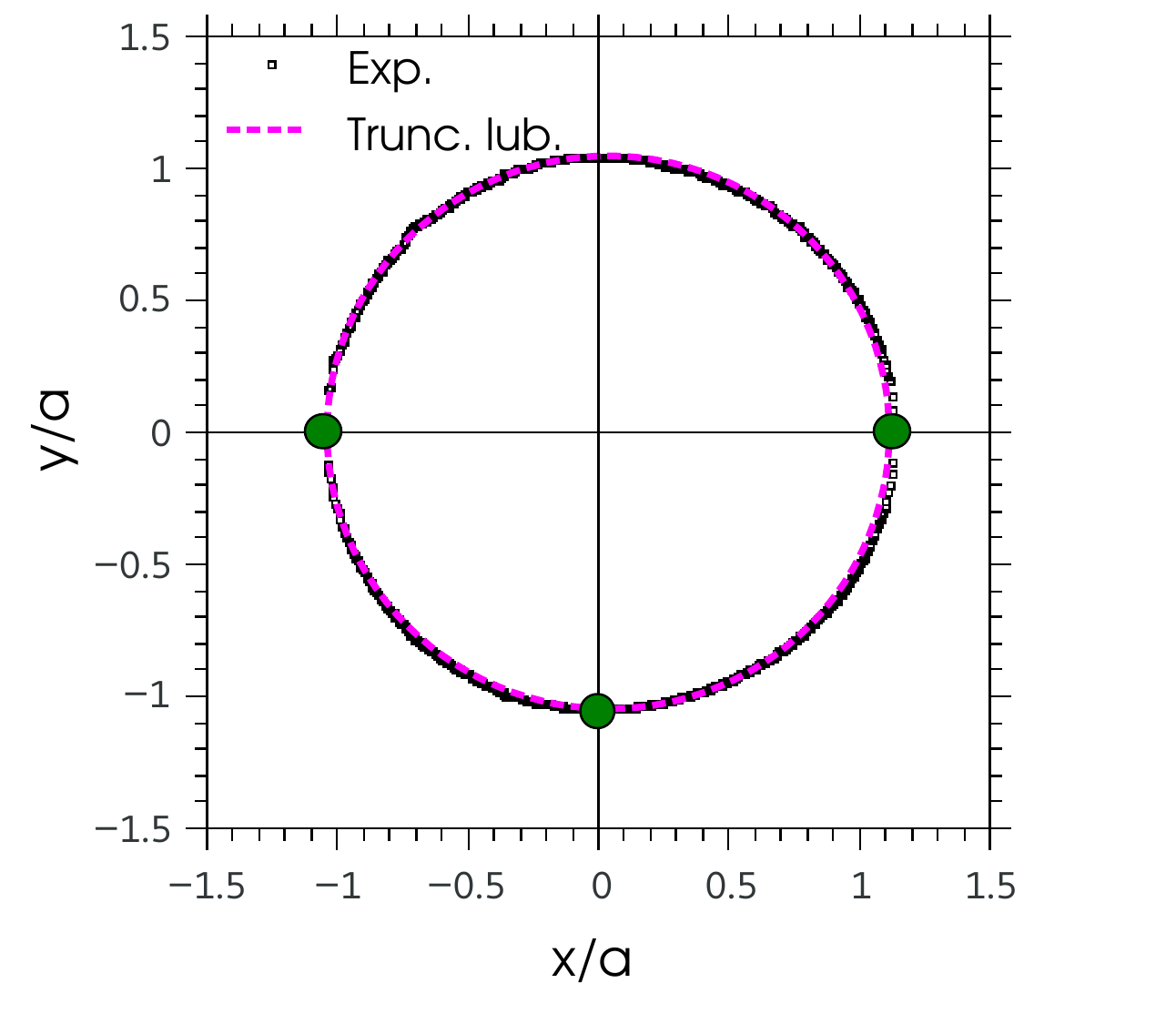}}
\subfigure[]{\includegraphics[width=0.329\textwidth]{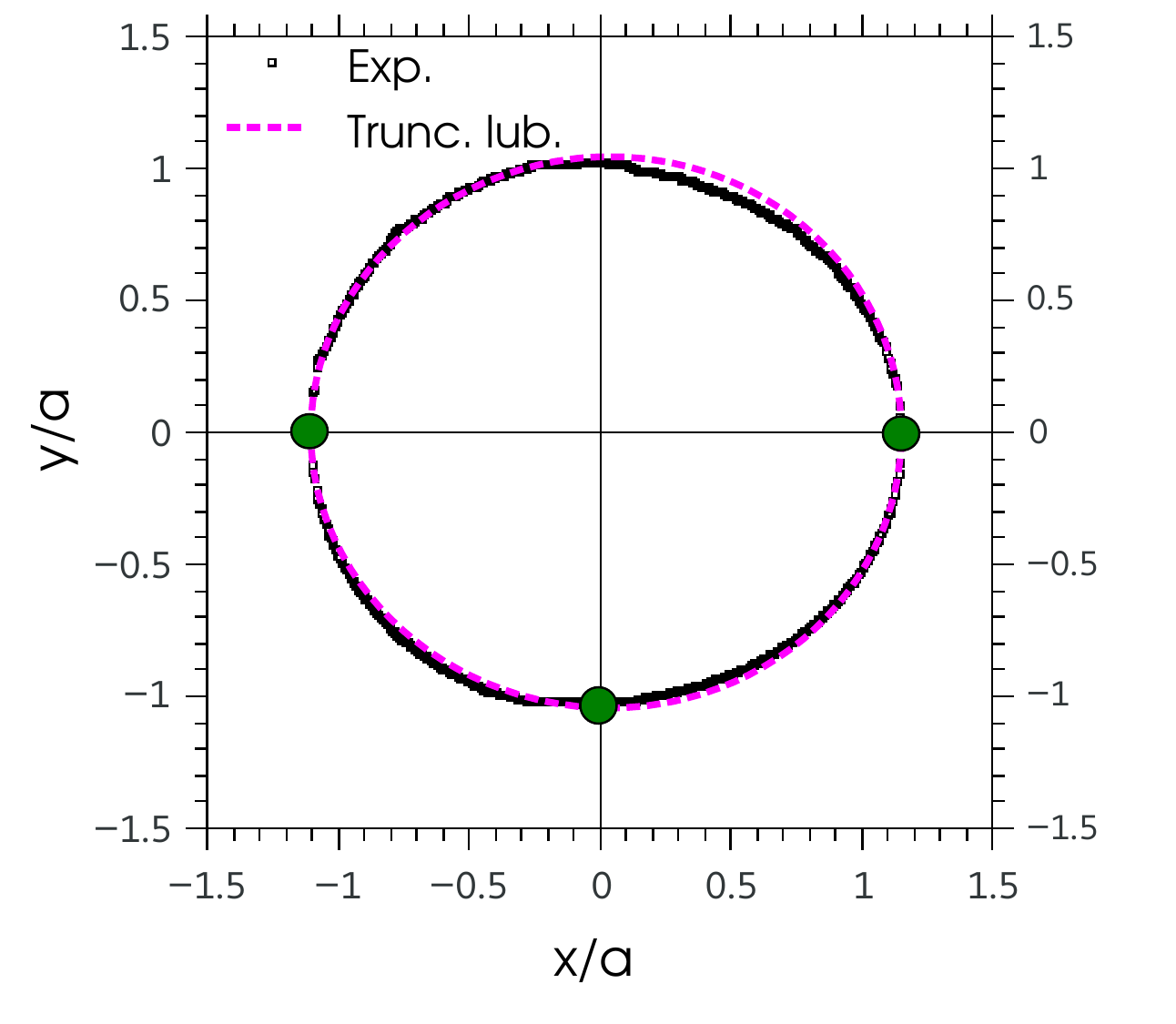}}
\caption{Comparison between the experimental footprint (symbols) and those predicted by the truncated approximation, Eq.~(\ref{eq:htrunc}) (dashed line), when using only the positions of the three points indicated by the green dots (and the experimental value $\theta_r$) for: (a) $\alpha=0^\circ$, (b) $\alpha=12.5^\circ$, and (c) $\alpha=25^\circ$.}
\label{fig:foot_trunc_exp}
\end{figure}

The determination of the constant $P$ is done by considering the contact angle
at $x_r$ as a known (measured) quantity, i.e. $\theta_r$. Thus, we have the
additional condition  
\begin{equation}
 \theta_r=  -\left. \frac{\partial h_{trunc}}{\partial r} \right|_{(r_r,0)},
\end{equation}
Finally, when all three constants $A_i$ and $P$ have been calculated, the
predicted footprint shape is obtained by the implicit equation
$h_{trunc}(r,\varphi)=0$. Figure~\ref{fig:foot_trunc_exp} shows comparisons of
footprints (dashed lines) with experimental data (symbols). For both $\alpha=0^\circ$ and 
$12.5^\circ$, the predicted footprints as given by
Eq.~(\ref{eq:htrunc}) are very close to the experimental data within the measurement error. The main difference is observed only on the right side for $\alpha=25^\circ$. However, we can say that there is a remarkable agreement with the experimental footprints.

\begin{figure}[htb]
\centering
\subfigure[]{\includegraphics[width=0.329\textwidth]{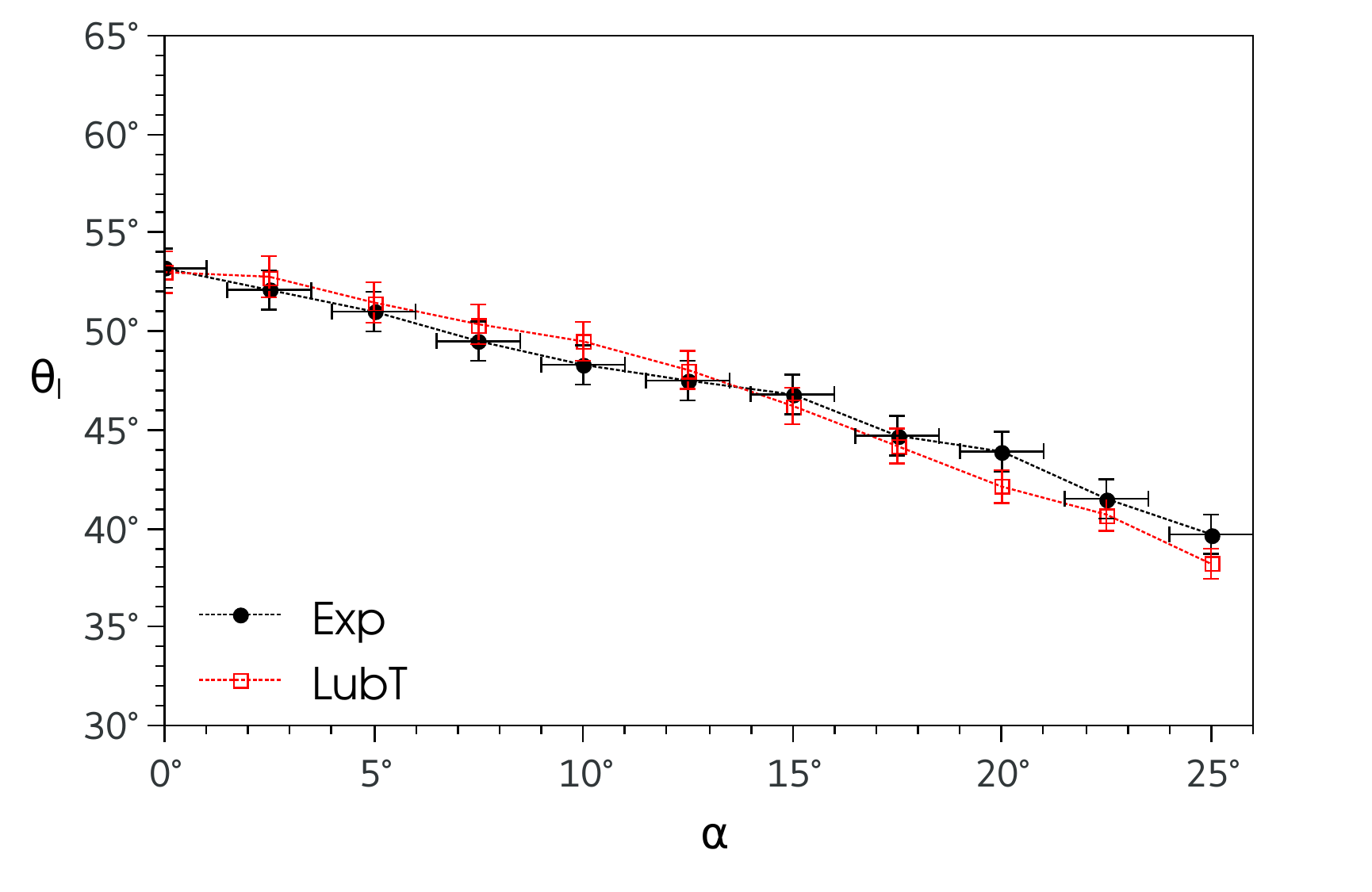}}
\subfigure[]{\includegraphics[width=0.329\textwidth]{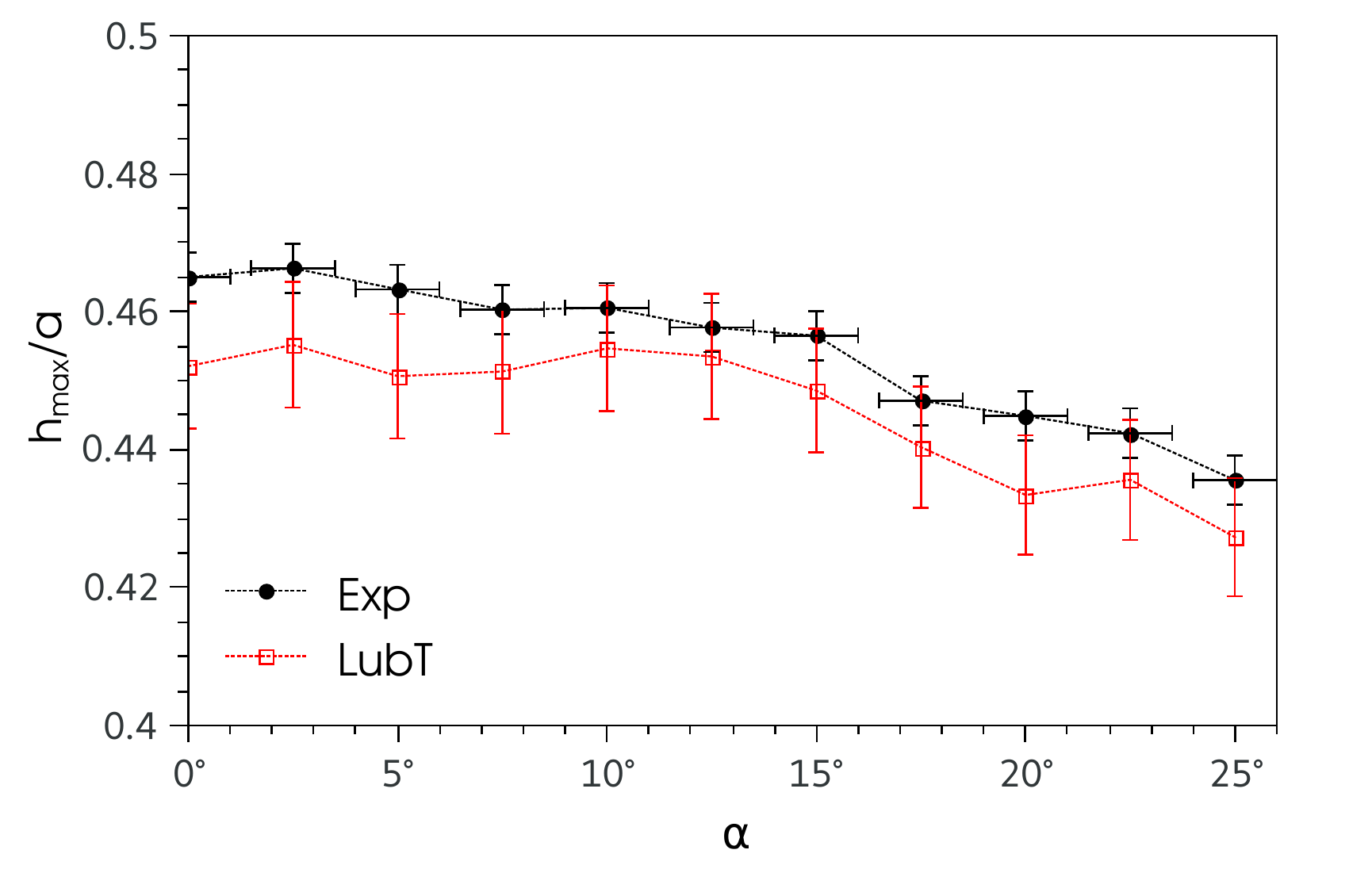}}
\subfigure[]{\includegraphics[width=0.329\textwidth]{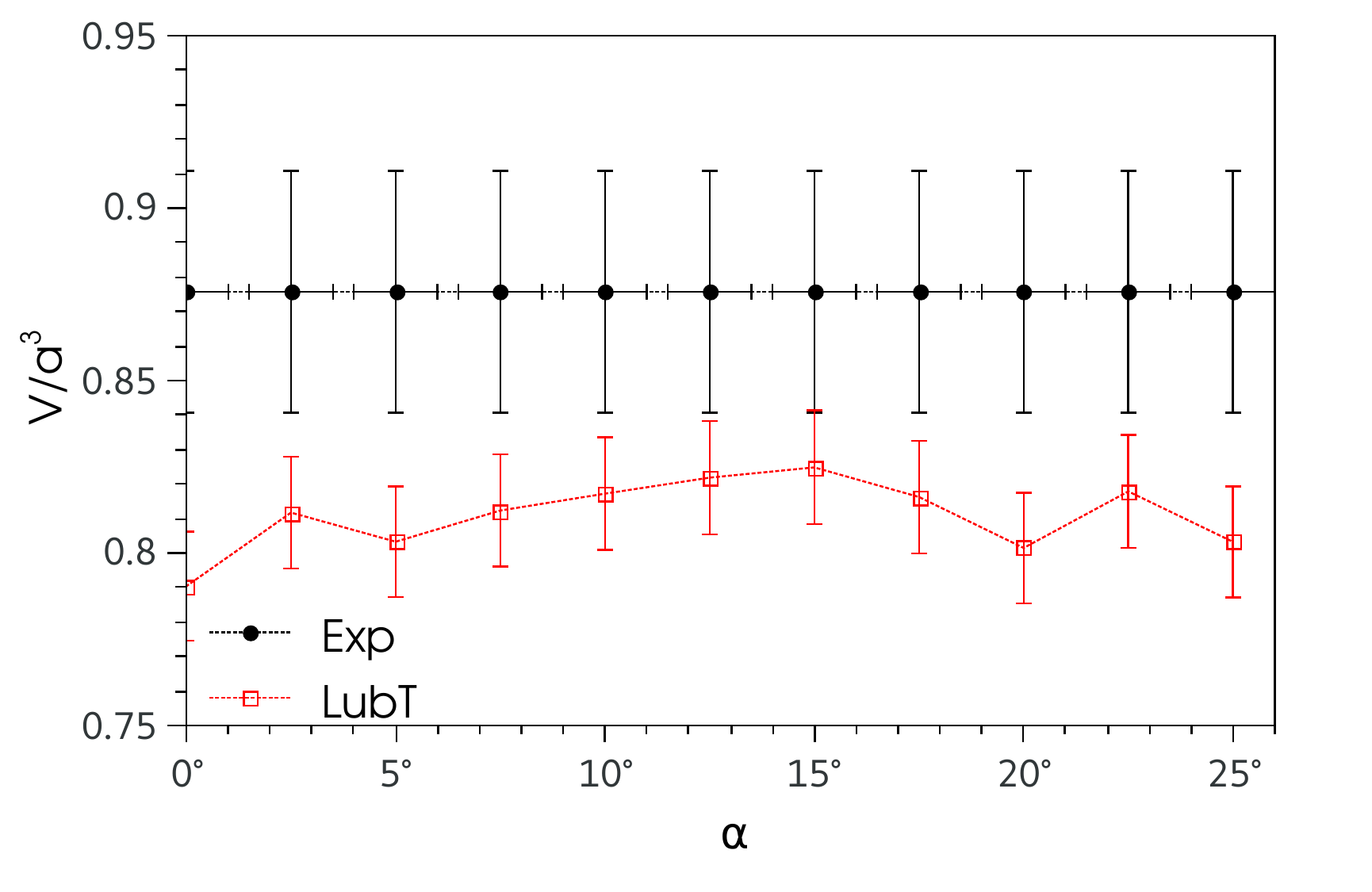}}
\caption{Comparison between the predictions of $h_{trunc}$ and the experimental
data for: (a) the contact angle at the left, b) the maximum drop thickness, and
(c) the drop volume.}
\label{fig:trunc_exp}
\end{figure}

The truncated solution also gives predictions for $\theta_l$, $h_{max}$ and $V$
(see Fig.~\ref{fig:trunc_exp}). For the contact angle in
Fig.~\ref{fig:trunc_exp} we use the expression 
\begin{equation}
 \theta_l=  \left. \frac{\partial h_{trunc}}{\partial r} \right|_{(r_l,\pi)}.
\end{equation}
Clearly, this solution improves the agreement with the experiments for
$\theta_l$ and $h_{max}$ respect to the numerical solutions with $\epsilon=0,1$
(see Fig.~\ref{fig:Exp_Num}a and c). However, the numerical solutions all have
the volume $V$ as given by the experiment, while now the predicted volume can
have a difference of at most $9\%$ respect to this value (see
Fig.~\ref{fig:trunc_exp}c). We conjecture that this smaller volume could yield
smaller values of both $\theta_l$ and $h_{max}$.

This analytical solution can also yield the azimuthal distribution of contact
angle, $\theta(\varphi)$. This is done by plugging Eq.~(\ref{eq:htrunc}) into
Eq.~(\ref{eq:thec_e0}) with the calculated values of the constants $A_i$ and
$P$. The predicted distributions for $\alpha=12.5^\circ$ and $25^\circ$ are
shown in Fig.~\ref{fig:the_num_exp} along with the numerical solutions for
$\epsilon=1$ and $\epsilon=0$. Interestingly, this approximation yields
intermediate values of $\theta$ for all $\varphi$.

\begin{figure}[htb]
\centering
\includegraphics[width=0.6\textwidth]{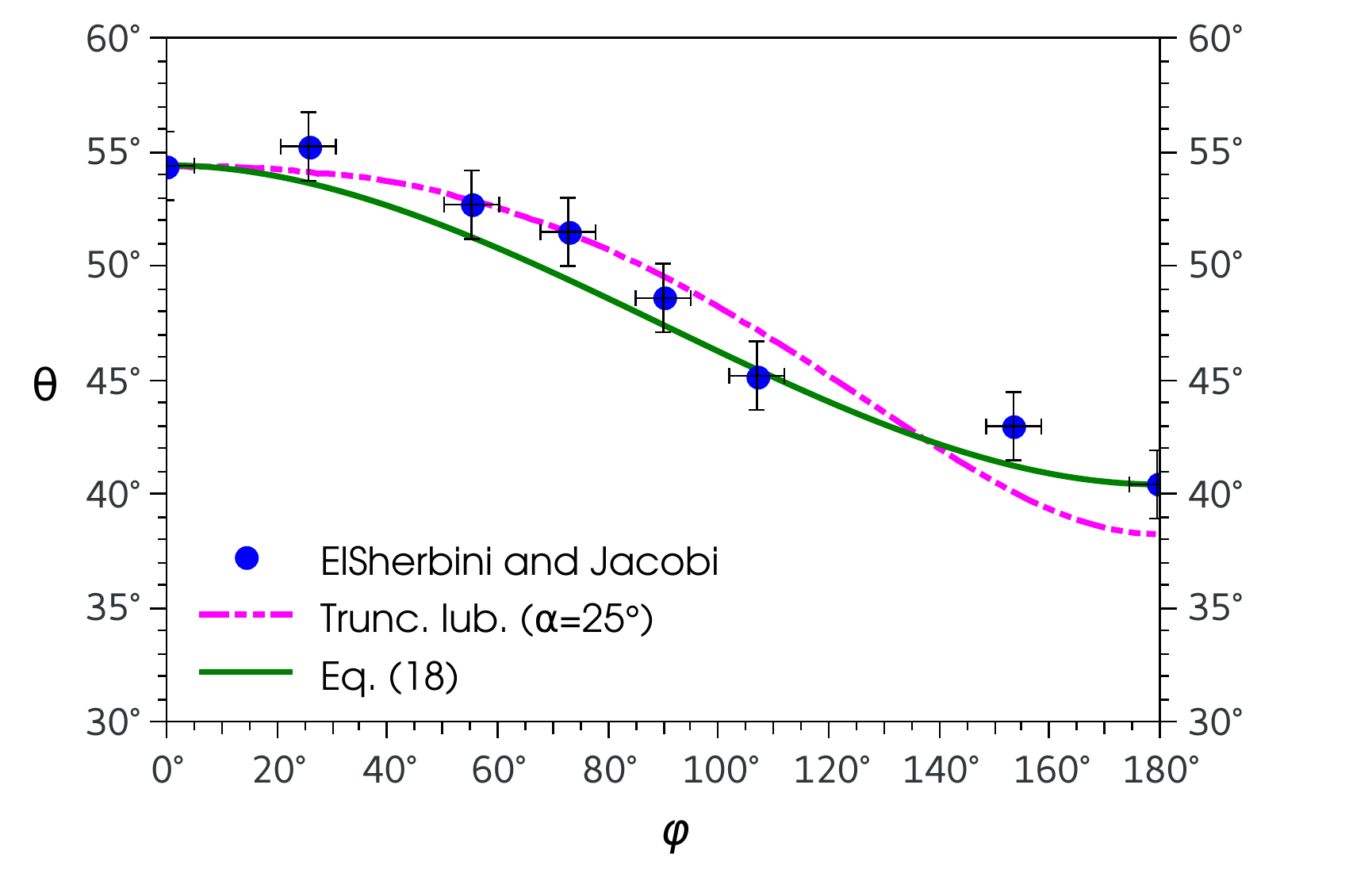}
\caption{Comparison between the experimental data of ElSherbini and Jacobi (blue circles with error bars; see Fig. 5a in~\cite{elsherbini_jcis04}) with $\theta(\varphi)$ as given by the solution of the truncated lubrication approximation (dashed line; see Fig.~\ref{fig:the_num_exp}b). The solid line corresponds to the fitting function given by Eq.~(\ref{eq:elsherb}).}
\label{fig:cangle_ES_LubT}
\end{figure}

In a further attempt to compare our theoretical prediction of $\theta(\varphi)$ with other works, we resort to the fitting cubic polynomial curve obtained by ElSherbini and Jacobi~\cite{elsherbini_jcis04}
\begin{equation}
 \theta(\varphi)= (\theta_r-\theta_l) \left( \frac{2 \varphi^3}{\pi^3} - \frac{3 \varphi^2}{\pi^2} + 1 \right) + \theta_l,
\label{eq:elsherb}
\end{equation}
where $\theta_r$ and $\theta_l$ are measured values. According to~\cite{elsherbini_jcis04}, this expression is valid for $10^\circ<\alpha<90^\circ$ and $0.75~mm^3<V<20~mm^3$. Fortunately, the values of $\theta_r$ and $\theta_l$ reported in their Fig. 5a are very close to ours for $\alpha=25^\circ$. Therefore, we compare in Fig.~\ref{fig:cangle_ES_LubT} our solution $\theta(\varphi)$ for the truncated lubrication approximation (see also Fig.~\ref{fig:the_num_exp}b) with their experimental data (full circles, whose error bars of $\pm 1.5^\circ$ and $\pm 5^\circ$ in $\theta$ and $\varphi$ are as explained in~\cite{elsherbini_jcis04}). Although the available data are limited to a single case ($\alpha=90^\circ$), we observe that our theory is in good agreement with them. Note that, while theoretical approach requires $\theta_r$ and predicts $\theta_l$, Eq.~(\ref{eq:elsherb}) needs an {\it a priori} knowledge of both values.

\section{Summary and conclusions}
\label{sec:conclu}

In this paper we report novel measurements of the contact angle hysteresis
curves and the possible equilibrium states by using the plane inclination,
$\alpha$, up to near a maximum value beyond which the drop starts sliding down.
The results are compared with those of usual hysteresis cycles obtained by
varying the drop volume, $V$, instead.The experiments show that both methods
yield hysteresis cycles with the same contact angles intervals, a factor
confirming that they are an intrinsic consequence of the wetting properties of
the system. 

While the drop footprint remains circular in the varying volume method, the
varying plane inclination method generates equilibrium drops with non--circular
footprints. This might seem a drawback of the method at first sight, but
provides interesting detailed insight on alternative equilibrium drop shapes
that are reached taking into account the hysteresis cycle. In particular, we
observe that the most important changes with respect to the initially circular
footprint for $\alpha=0$ occur in the downhill region (see e.g.
Fig.~\ref{fig:hxFoot}b). However, these deformations are not the same at
different parts of the cycle, even though they correspond to equal $\alpha$ (see
Fig.~\ref{fig:hx_alpha125_R12}b). On the other hand, the present experimental
results have been compared successfully with previously reported data (see
Fig.~\ref{fig:ratio_Theta}).

In order to describe the drop shape (both free surface and footprint) at
different inclination angles, we have solved the pressure equilibrium equation
out of and within the approximation of small contact angles. For this goal, we
have resorted to numerical solutions of both the full and approximated
equations, as well as to the analytical solution of the latter in the form of a
series. The numerical task has shown quantitatively how good are the approximations 
based on the long--wave theory. As a result, it turns out that they are
good enough, since their differences with the measured parameters are within the
experimental error interval.

One drawback of these numerical and analytical solutions is that the complete
determination of the drop shape requires, for instance, to have measured
previously the full shape of the footprint. However, we prove that a truncated
expression of the analytical solution suffices to determine both the free
surface and the full footprint from a very small set of data at the footprint,
namely: the positions of three points on the curve, and a contact angle at one
of them. Our theoretical solution has practical use, since two of those points
as well as the contact angle are easily measured from the lateral view of the
drop, which is the usual setup in most of the experiments reported in the
literature. The third point can be extracted from a knowledge of the transverse
width of the drop.

The contact angle variation around the drop periphery, $\theta(\varphi)$, is
another important characteristic of the drop shape that can be extracted from
the numerical and theoretical analysis presented here. We should note that this
angular information is very difficult to measure, specially for an inclined
plane. The implementation of a refractive technique similar to that used in~\cite{gonzalez_07,rava_pof16}, which yields $\theta$ all along the drop periphery, is left for future work. Up to our knowledge, theoretical models have been developed in the
literature only for small inclinations and small drop volumes, when the
footprint can be considered circular. We report numerical calculations of
$\theta(\varphi)$ without any restriction on $\alpha$ or $V$, and we also obtain
this curve from the theoretical solution with the truncated series. When the former solution is applied by using the measured value of $\theta_r$ ($\varphi=0$), a very good prediction of $\theta_l$ ($\varphi=\pi$) is obtained. Unfortunately, comparisons of $\theta(\varphi)$ with experimental data for at other values of $\varphi$ is not possible, since these measurements have not yet been reported. 

In summary, we develope an alternative method to measure the static hysteresis cycle that yields non trivial equilibrium drop shapes, whose main features have been studied here.

\acknowledgments
P. Ravazzoli and I. Cuellar acknowledge postgraduate student fellowships from Consejo Nacional de Investigaciones Cient{\'i}ficas y T{\'e}cnicas (CONICET, Argentina) with Res. 1202/2014 and Res. 4209/2017, repectively. J. Diez and A. Gonz{\'a}lez acknowledge support from Agencia Nacional de Promoci{\'o}n Cient{\'i}fica y Tecnol{\'o}gica (ANPCyT, Argentina) with Grant No. PICT 1067/2016.

\end{document}